%% file: main.tex
\documentclass[twocolumn,apj,iop]{openjournal}

\usepackage{amsmath}
\usepackage{lineno}
\usepackage{color}
\usepackage{url}
\usepackage{hyperref}
\hypersetup{colorlinks=true,linkcolor=blue,citecolor=blue,filecolor=blue,urlcolor=blue}
\usepackage{cleveref}
\usepackage{flafter} 
\usepackage{float}
\usepackage{longtable}

\newcommand{\msun}{\ensuremath{M_\odot/h_{70} }}
\newcommand{\mass}{\mbox{$M_{\mbox{\scriptsize 500c}}$}}

\newcommand{\spitzer}{{\sl Spitzer}}
\newcommand{\planck}{{\sl Planck}}

\newcommand{\nSN}{\ensuremath{\xi}}

\newcommand{\sqdeg}{deg$^2$}
\newcommand{\um}{\ensuremath{\mu{\rm m}} }

\newcommand{\beq}{\begin{equation}}
\newcommand{\eeq}{\end{equation}}
\newcommand{\beqar}{\begin{eqnarray}}
\newcommand{\eeqar}{\end{eqnarray}}

\newcommand{\ncand}{689}
\newcommand{\nconfirm}{544} 

\usepackage{perpage}
\MakePerPage{footnote}

\graphicspath{{figures/}}

\begin{document}

\begin{nolinenumbers}
\vspace*{-\headsep}\vspace*{\headheight}
\footnotesize \hfill FERMILAB-PUB-23-571-PPD\\
\vspace*{-\headsep}\vspace*{\headheight}
\footnotesize \hfill DES-2023-0788
\end{nolinenumbers}

\title{Galaxy Clusters Discovered via the Thermal Sunyaev-Zel'dovich Effect in the 500-square-degree SPTpol Survey\vspace{-1.25cm}}

\input{500d_authors}

\shorttitle{Galaxy Clusters Discovered in the SPTpol 500d Survey}
\shortauthors{Bleem et al.}

\thanks{$^{\star}$E-mail:lbleem@anl.gov}

\begin{abstract}
We present a catalog of  \ncand \ galaxy cluster candidates detected at significance $\xi>4$ via their thermal Sunyaev-Zel'dovich (SZ) effect signature in 95 and 150 GHz data from the 500-square-degree SPTpol survey. 
 We use optical and infrared data from the Dark Energy Camera and the Wide-field Infrared Survey Explorer (WISE) and \spitzer \ satellites, to confirm \nconfirm \ of these candidates as clusters with $\sim94\%$ purity. 
The sample has an approximately redshift-independent mass threshold at redshift $z>0.25$.  The 
confirmed sample spans  $1.5  \times 10^{14} < M_{500c} < 9 \times 10^{14} $\msun \  and $0.03<z\lesssim1.6$  in mass and redshift, respectively, with a median mass of  $2.5 \times 10^{14} $\msun \ and median redshift $z=0.7$; 21\% of the confirmed clusters are at $z>1$.
We use external radio data from the Sydney University Molonglo Sky Survey (SUMSS) to estimate contamination to the SZ signal from synchrotron sources. 
The contamination reduces the recovered $\xi$  by a median value of 0.032, or $\sim0.8\%$ of the $\xi=4$ threshold value, and $\sim7\%$ of candidates have a predicted contamination greater than $\Delta \xi = 1$.
With the exception of a small number of systems $(<1\%)$,  an analysis of clusters detected in single-frequency 95 and 150 GHz data shows no significant contamination of the SZ signal by emission from dusty or synchrotron sources. 
This cluster sample, representing the deepest SZ-selected cluster sample to-date, will be a key component in upcoming astrophysical and cosmological analyses of clusters.  In addition to the cluster catalog, we also release the millimeter-wave maps and associated data products used to produce this sample. These maps have depths of 5.3 (11.7) $\mu$K$_{\textrm{CMB}}$-arcmin at 150 (95) GHz and an effective angular resolution of 1$\farcm{2}$ (1$\farcm{7}$). The SPTpol products are available  at \url{https://pole.uchicago.edu/public/data/sptpol_500d_clusters/index.html}, and the NASA LAMBDA website. An interactive sky server with the SPTpol maps and Dark Energy Survey data release 2 images is also available at NCSA \url{https://skyviewer.ncsa.illinois.edu}. 
\end{abstract}


\section{Introduction} \label{sec:intro}

The advent of high-resolution cosmic microwave background (CMB) surveys has revolutionized the field of galaxy cluster science by enabling the identification of approximately mass-limited samples of massive galaxy clusters \citep{vanderlinde10,bleem15b,planck15-27,hilton18,bleem20} via the thermal Sunyaev-Zel’dovich (SZ) effect \citep{sunyaev72}. 
These SZ samples have been used to place stringent constraints on cosmological models \citep{planck15-24,bocquet19,zubeldia19,salvati22}, the evolution of the intracluster medium  \citep{mcdonald17,ghirardini21,flores21,chexmate21,anbajagane22,ruppin22,olivares22}  and the evolution of galaxies residing in clusters \citep{hennig17,chiu18,strazzullo19,khullar22,kim23,somboonpanyakul22}.
The next generation of CMB surveys, with significantly increased sensitivity, is continuing this revolution by probing lower-mass and correspondingly higher-redshift systems \citep{huang20,hilton21}.  

With this increased sensitivity comes both new opportunities \citep{simonsobservatorycollab19,chaubal22} and challenges \citep{melin18, zubeldia23} for cosmological and astrophysical analyses with SZ clusters.  
The enlarged sample sizes will also strengthen cross-wavelength analyses, for which comparisons of cluster samples selected via different techniques provide powerful ways to validate the robustness of analysis inferences and can signal the presence of unmitigated systematic errors \citep{grandis21,costanzi21, orlowski21}.

In this work we continue the push to discover SZ clusters in low-noise and high-resolution CMB survey data by presenting a galaxy cluster sample constructed using the 500-square-degree SPTpol survey.
This survey,  which covers five times the area used in the 100-square-degree SPTpol field (hereafter SPTpol 100d) cluster analysis  \citep{huang20}, benefits from approximately two-to-three times lower noise than the survey data used to construct the published cluster samples from SPT-SZ \citep{bleem15b} and ACTpol \citep{hilton21}.
 The new cluster sample is composed of \ncand \ cluster candidates detected at significance $\xi>4$, and we have confirmed \nconfirm  \ of these candidates as galaxy clusters using optical and infrared imaging data.   
Here we provide the complete cluster candidate list and---where available---redshifts, estimated masses, and select optical properties. 

We organize our description of the identification and characterization of this new sample as follows. 
In Section \ref{sec:data} we describe the collection of the millimeter-wave data and its processing into the maps used for cluster identification.  
We detail the procedure by which we identify SZ cluster candidates in these maps in  
Section \ref{sec:clusterfinding}. 
In Section \ref{sec:followup} we describe the optical and infrared data sets we use to confirm candidates as galaxy clusters by the techniques described in Section \ref{sec:confirmation}. 
In Section \ref{sec:mass} we describe how we estimate cluster masses and estimate the purity of our cluster sample.
In Section \ref{sec:biases} we describe a number of systematic checks to explore possible astrophysical contamination of the SZ signal.
In Section \ref{sec:catalog} we present the properties of the resulting SZ cluster catalog  and  compare select properties of this catalog to other SZ- and optically selected cluster samples. 
Finally, we summarize our results and conclusions in Section  \ref{sec:conclusion}.

In addition to the cluster sample we also release the  SPTpol maps and associated data products used in the construction of this catalog.  
These products can be found on both the NASA Legacy Archive for Microwave Background Data (LAMBDA)\footnote{\url{https://lambda.gsfc.nasa.gov/product/spt/sptpol\_prod\_table.html}} and the South Pole Telescope (SPT) collaboration's website\footnote{\url{https://pole.uchicago.edu/public/Data\%20Releases.html}}. 
Sky maps for the 500d region from SPTpol single-frequency data as well as images from the Dark Energy Survey (DES)-DR2 \citep{abbott21} dataset can be accessed for exploration at the National Center for Supercomputing Applications (NCSA) Skyviewer.\footnote{ \url{https://skyviewer.ncsa.illinois.edu}}  

Finally we note that, where applicable, we assume a fiducial $\Lambda$CDM cosmology with $\sigma_8=0.80$, $\Omega_b = 0.046$, $\Omega_m = 0.30$, $h = 0.70$, $n_s(k_s=0.002) = 0.972$, and $\Sigma m_\nu=0.06$ eV.  
Magnitudes are reported in the AB system \citep{oke74}.
Cluster masses are reported in terms of  $M_{500c}$, which is defined as the mass enclosed within a radius, $r_{500c}$, at which the average enclosed density is 500$\times$ the critical density at the cluster redshift.

\begin{figure*}[t]
\begin{center}
\includegraphics[width=6.8in]{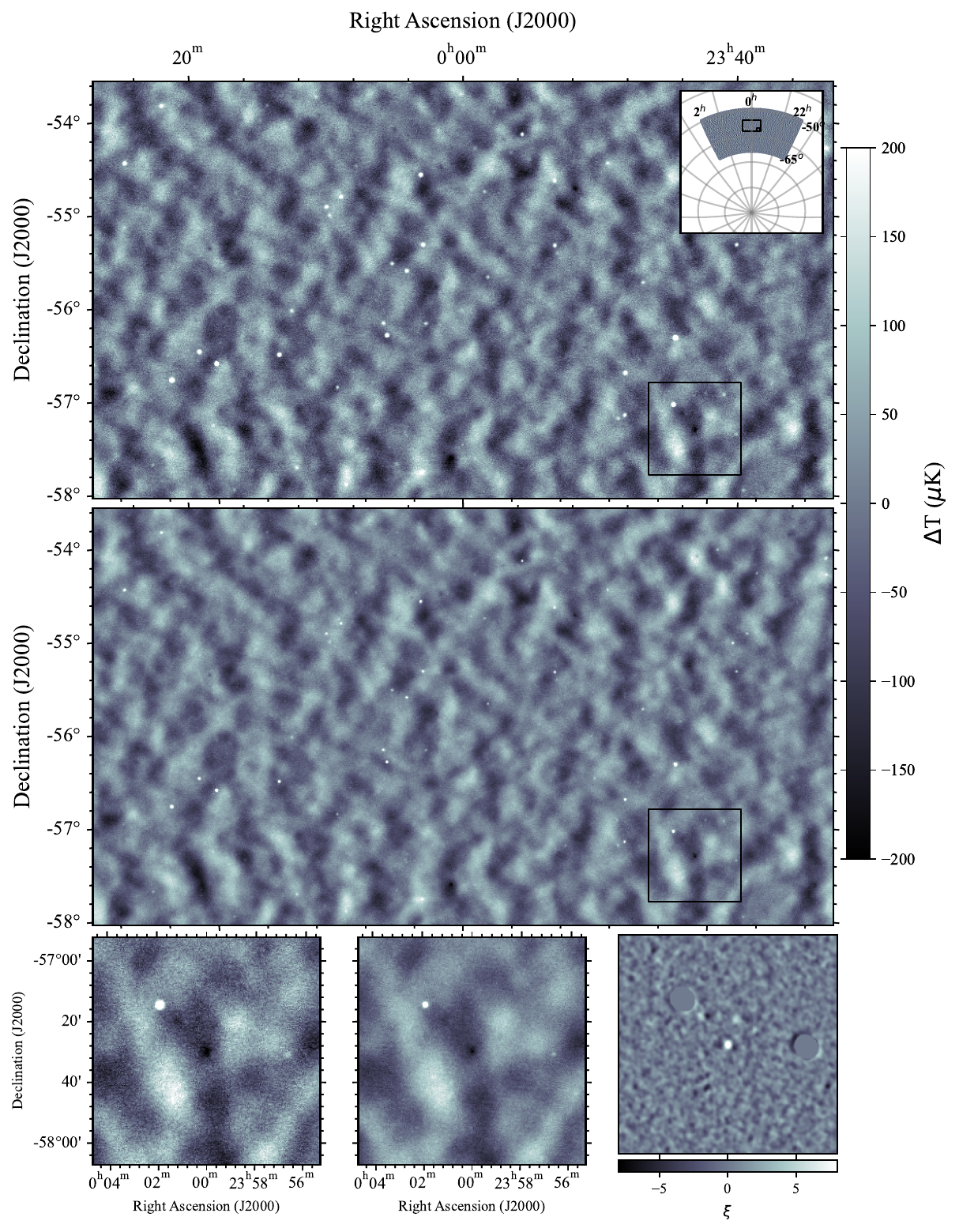}
\caption{Sequentially zoomed-in images of the SPTpol 500d field, with the full field shown in inset at top. \textit{(Top panel)}  24 deg$^2$ of 95~GHz data \textit{(Middle panel)} The same region at 150~GHz.  The full field temperature maps and associated data products are released along with this work.  \textit{(Bottom panel)} To better illustrate the sensitivity of these maps to small scales, we  further zoom in to a 1.25 $\times$ 1.25 degree  region (outlined in black in above panels) centered on SPT-CL~J2341-5724, a $\xi=14.3$ cluster at $z=1.259$. Plotted from left-to-right are the 95 and 150 GHz temperature maps and the $\xi$-map constructed for optimal detection of clusters with $\theta_\textrm{core}=0.25$. Sources brighter than 6 mJy at 150~GHz were masked in the cluster detection step, with two examples visible in the bottom right panel. }
\label{fig:fieldplot}
\end{center}
\end{figure*}

\section{Observations and Data Reduction}\label{sec:data}
The cluster catalog presented in this work was constructed via an analysis of 95 and 150~GHz data from the primary SPTpol   \citep{austermann12}  survey field. 
This 500-square-degree field is centered at right ascension and declination ($\alpha$=0$^{h}$, $\delta$= $-$57.5$^{\circ}$) and is hereafter referred to as the SPTpol 500d field; see Figure \ref{fig:fieldplot}.  
The maps have depths of 11.7 and 5.3 $\mu$K-arcmin at 95 and 150 GHz, respectively, and effective resolutions of 1$\farcm{7}$ and 1$\farcm{2}$ set by the instrumental response (i.e., telescope beam). 
The SPTpol data in this field have been the focus of a number of previous analyses including characterizations of low- and high-$\ell$ CMB temperature and polarization power spectra \citep{henning18,sayre20, reichardt21}, the CMB gravitational lensing signal \citep{wu19, bicepkeckspt21, raghunathan19b}, and the properties of emissive sources \citep{gupta19}. 
The overall map-making and data processing procedures utilized here closely follow those of previous efforts. 
In this section we provide a brief summary of these procedures, highlighting changes specific to this work, and refer readers to prior publications for more details.

\subsection{Data Processing}\label{subsec:processing}
The maps presented here are the weighted sum of $>4000$ individual observations ($\sim$10,200 hours) of the SPTpol 500d field acquired between 2013 April 30 and 2016 Aug 20. 
The majority of the observations were obtained by scanning the telescope in azimuth back and forth across the field, stepping in elevation, and repeating this process until the full field was covered.  As detailed in \citet{henning18},  most of the data obtained before 2014 May 29 was acquired  instead in ``lead-trail" mode. 
In this mode, the field is split into two equal halves via an equal division in right ascension. 
These two subfields were then sequentially scanned, with starting times offset owing to sky rotation, such that the same azimuthal range was covered by the telescope in each subfield.
For this analysis we combine these ``lead" and ``trail" observations into full-field observations to match the rest of the 500d dataset. 

The SPTpol data processing pipeline converts time-ordered electrical signals recorded by the camera into calibrated maps of the millimeter-wave (mm-wave) sky. 
We apply standard detector quality cuts \citep{crites15} and electrical cross talk corrections to the time-ordered data (TOD). 
Following \citet{bleem20}, to minimize the impact of low-frequency noise from the atmosphere and detector readout, both a common mode filter that removes the mean of all detectors in each frequency band and a seventh-order Legendre polynomial were fit to and subtracted from the TOD for each azimuthal scan.  
A scan-direction high-pass filter at angular multipole $\ell=300$ and similar low-pass filter at $\ell=20,000$ were then applied.  
Bright emissive sources detected at 150 GHz at $>6$ mJy were masked with masks of 4$\arcmin$ radii during these filtering steps so as to not bias the fits or imprint artifacts in the resulting maps.

Using the telescope pointing model and weights for the individual bolometers based on their noise in the 1-3 Hz band, the TOD was binned into $0\farcm{25}$ pixels in the Sanson-Flamsteed projection \citep{calabretta02}.  
The noise properties of each single-observation map were characterized and, following removal of a small number of maps with anomalous noise behavior,  the observations were then combined via inverse noise-variance weighting to produce the final coadded maps of the field.    

\subsection{Removal of Emissive Sources that Cause Spurious Cluster Candidates}\label{subsec:sourcesub}

As discussed in \citet{huang20}, unmasked emissive sources in the maps can lead to spurious cluster candidates owing to decrements produced by the sources ``ringing" when the maps are high-pass filtered. 
To mitigate the number of such spurious detections, we identify and remove moderate signal-to-noise (S/N) sources below our masking threshold from the maps before cluster detection occurs. 

We first filter each individual frequency map with a filter optimized to detect point sources and construct a catalog of emissive sources detected at S/N $>6$ and below the masking threshold of 6 mJy at 150 GHz used in the map construction. 
This corresponds to thresholds of $\sim$ 4 mJy (3.5 mJy) at 95 GHz (150 GHz). 
A total of 348 sources were detected at 95~GHz and  382 at 150~GHz (245 at S/N$>6$ in both maps). 
When sources were only detected in one of the two frequency maps the other frequency map was forced-photometered to recover the missing flux. 

\begin{figure}
\begin{center}
\includegraphics[width=3in]{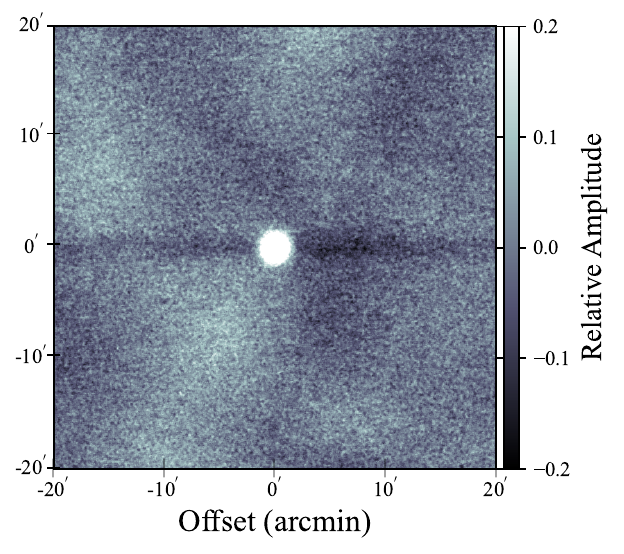}
\caption{40$\arcmin \times 40\arcmin$ normalized median stack of sources at 95 GHz used to make the source subtraction template; a similar template was constructed at 150 GHz.  The negative filtering artifacts along the scan (horizontal) direction are clearly visible even in these stacks which have significant residual large scale noise contributions from the primary CMB.} \label{fig:sourcestack}
\end{center}
\end{figure}

An empirical template of the 2D source profile, incorporating the effects of the beam and transfer function, was constructed by creating a flux-normalized median stack of the raw unfiltered maps at the locations of the $\sim$~200 brightest sources at each frequency (see Figure \ref{fig:sourcestack}). 
A flux-scaled copy of this template was then subtracted from the raw maps at the location of each of the detected sources. 
Analysis of the cluster catalog (whose construction is detailed in the next section) showed this simple procedure was sufficient to remove the spurious candidates  associated with these  emissive sources. 
Artifacts from remaining lower flux sources contribute $<2\sigma$ SZ-like signals in the cluster detection maps. 
We use these source subtracted maps in all of our cluster identification steps.

\section{Identification of Cluster Candidates}\label{sec:clusterfinding}

We closely follow the procedure utilized in previous SPT publications  \citep[most recently][]{huang20, bleem20} to identify cluster candidates in the SPTpol 500d field with some small changes we explain below.  
As in these works, we filter the maps with a spatial-spectral filter that has been optimized to isolate cluster signals in the presence of frequency- and scale-dependent sources of noise \citep{melin06}.
The spatial profile of clusters are modeled as a series of projected spherical $\beta$ profiles \citep{cavaliere76}
\begin{equation}
\Delta T = \Delta T_{0}(1+ \theta^{2}/\theta_\mathrm{c}^{2})^{-(3\beta-1)/2}
\label{eqn:beta}
\end{equation}
with $\beta=1$, a free normalization, $\Delta T_{0}$, and with core radii, $\theta_\mathrm{c}$, that are allowed to vary in twelve equally spaced steps from $0\farcm25$ to $3\arcmin$. 
The frequency dependence of the temperature signal in each map, $\Delta T$, is given by the thermal SZ effect \citep{sunyaev72}. 
At the SPTpol effective band centers of 95.9 and 148.5 GHz for a non-relativistic thermal SZ spectrum, this results in the thermal SZ signal being $1.6\times$ brighter  (in CMB temperature units) in the 95 GHz than 150 GHz data.  This enhancement of the signal amplitude largely offsets the higher instrumental noise in the 95 GHz data, and---combined with the impacts of astrophysical noise discussed below---results in the two channels contributing close to equal weight in the construction of the cluster catalog. 
The noise in the SPTpol maps is broadly composed of two components: (1) astrophysical/cosmological ``noise" arising from fluctuations in the primary CMB and emission from extragalactic sources and (2) noise specific to our observations arising from the SPTpol instrument as well as residual atmospheric contamination.  

The SPTpol data are deep enough that---for identical reasons as in the case of moderate signal-to-noise emissive sources discussed in  Section \ref{subsec:sourcesub}---the Fourier filtering induces spurious candidates around massive clusters along the telescope scan direction.   
As the sky density of such clusters is significantly lower than that of the emissive sources---0.09 per deg$^{2}$ at the conservative threshold we apply below---and they have a less well defined shape, we do not attempt to subtract a model of these clusters from the map. 
Instead we use a two-step procedure to construct a clean cluster candidate list that also recovers the properties of the most massive systems.

In the first step, we perform a cluster detection run on the maps masking both emissive sources with flux  {$>6$~mJy} at 150 GHz and massive clusters previously detected in the SPT-SZ survey\footnote{The SPT-SZ survey covers all but 1 \sqdeg \ of the SPTpol 500d field.} at significance\footnote{In this work, as in all previous SPT cluster publications, we define  significance, $\xi$, as the maximum detection significance across the 12 matched filter scales.} $\xi \ge$ 6. 
This corresponds to a masking threshold  of \mass \  $\gtrsim 4.5 \times 10^{14} \msun$.  
We use 4\arcmin \ radii masks and further exclude cluster candidates detected within 8$\arcmin$ of these objects. 
Cluster candidates identified with $\xi>4$ in this step from the SPTpol data form the core of the new cluster sample. 

In the second step, to include the massive clusters masked in the previous step in our catalog, we perform a second filtering of the maps in which only the emissive sources  $>6$ mJy at 150 GHz are masked. 
We then add only the new detections of previously masked clusters and their properties to our cluster candidate list. 
There were no clusters masked in the first step that did not exceed our $\xi \ge 4$ cut.
We validated through comparison of the noise properties in the two filtering steps that the presence of the most massive systems does not impact the estimated noise.  We additionally visually  inspected cutouts of the raw and filtered maps around all $\xi>4$ cluster candidates to validate our treatment of both emissive sources and the highest signficance clusters. 
In total,  after all masking is accounted for, 460 of the 498 \sqdeg \ uniform depth region (92\%) of the SPTpol 500d field was included in the cluster search.

\section{Follow-up Observations}\label{sec:followup}
SZ galaxy cluster candidates are confirmed and redshifts obtained via the identification of significant galaxy overdensities in optical and/or infrared imaging data. 
These data are drawn from both wide-field imaging surveys and targeted photometric and spectroscopic follow-up observations.
The majority of our candidates are confirmed using data from the DES \citep{flaugher15} and the highest-redshift clusters are confirmed using observations from the Wide-field Infrared Survey Explorer \citep[WISE;][]{wright10} and \spitzer/IRAC \citep{fazio04}. 
A small number of candidates at the borders of the SPTpol survey do not overlap with DES; these candidates are characterized using data from the DECam Legacy Survey \citep[DECaLS;][]{dey19}.

\subsection{Optical/Near-Infrared Imaging from the Dark Energy Camera}
The DES is a $\sim$5000 \sqdeg \ 5 band \textit{grizY} optical and near-infrared imaging survey that was conducted using the 4 m Blanco telescope at Cerro Tololo Inter-American Observatory in Chile. 
Data were acquired over 760 nights between August 2013 and January 2019 \citep{abbott21}.  
In this analysis, we make use of photometric catalogs extracted from the full 6 year coadds. 
These data reach median 10$\sigma$ coadded depths in 1\farcs{95} apertures of [24.7, 24.4, 23.8, 23.1, 21.7] magnitude in the \textit{grizY} bands. 
The DES data cover almost the entire SPTpol 500d field and are typically deep enough to robustly confirm SPTpol cluster candidates to redshift \textit{z}$\sim1.1$.

Eighteen candidates at the edges of the SPTpol 500d field fall outside the DES coverage region. 
We use \textit{griz} photometry data from the tenth data release of  DECaLS\footnote{\url{https://www.legacysurvey.org/dr10/description/}} to confirm and obtain redshifts for 11 of these candidates; the remaining 7 candidates are unconfirmed after optical/infrared analysis.  
Tests of cluster redshift estimates using systems in regions where DES and DECaLS overlap show good agreement.

\subsection{Wide Field Infrared Explorer} 
To detect candidates at higher redshifts, we analyze data from the all-sky WISE dataset. 
Specifically  we use  3.4 $\mu$m and 4.6 $\mu$m data ({\sl W1} and {\sl W2} band, respectively) from the ``unWISE''  analysis of \citet{schlafly19} 
which combined five years of data from WISE and NEOWISE \citep{mainzer14} data at native WISE resolution.   

\subsection{Spitzer/IRAC}
A subset of high-redshift cluster candidates were also targeted with \spitzer/IRAC at 3.6 and 4.5 $\mu$m ([3.6] and [4.5] bands). 
These deeper data are significantly higher resolution than WISE ([3.6]$_\textrm{FWHM}\sim2\arcsec$ compared to 6$\arcsec$ from WISE {\sl W1}), allowing us to confirm additional high-redshift clusters as well as to validate our WISE analysis. 
\spitzer/IRAC observations  of cluster candidates are drawn from three sources: 50 candidates were previously targeted for follow-up observations in the production of the SPT-SZ cluster sample \citep[PI: Brodwin, see details in][]{bleem15b}, 84 cluster candidates were imaged in new targeted observations in \spitzer  \ cycles 11, 12, and 14 (Program IDs 11096, 12073, 14096; PI: Bleem), and 163 cluster candidates (at any redshift) were within the footprint of the \spitzer \ South Pole Telescope Deep Field \citep[SSDF;][]{ashby13}, with 31 of these candidates also having deeper targeted follow-up from the earlier SPT-SZ work. 

The SPTpol candidates imaged in Cycles 11-14 were initially detected in preliminary versions of the SPTpol 500d cluster sample, and were selected for infrared follow-up based on the absence of galaxy overdensity counterparts in then existing DES data \citep[DES Year 1 and DES Year 3, depending on the cycle;][]{abbott18b}. 
In the new \spitzer \ observations, candidates were  observed in each band in 6 $\times$ 30~s exposures and the data were reduced following the procedures detailed in  \citet{ashby09}.  
This exposure time is sufficient to reach a 5$\sigma$ sensitivity of 4.8$\mu$Jy  at 3.6$\mu$m in an aperture-corrected 4\arcsec\
aperture. 
Here we use 4$\arcsec$ diameter aperture-corrected magnitudes.

\section{Cluster confirmation and Redshift Estimation}\label{sec:confirmation}
The majority of newly confirmed cluster candidates in this work were confirmed using the multi-component matched filter cluster confirmation tool (MCMF) previously described in detail in \citet{klein18,  klein19}.

\subsection{MCMF}\label{sec:mcmf}

As explained in \citet{klein18, klein19}, the MCMF algorithm was designed to provide---amongst other properties---cluster confirmation, redshift, and cluster galaxy richness information for samples of X-ray and SZ clusters by robustly identifying associated cluster galaxy counterparts.
We provide a brief overview of the method here, including choices specific to this analysis, and refer readers to  previous works for more details. 
 The MCMF algorithm works as follows: 
\begin{itemize}
\item At the location of each cluster candidate, local background-corrected cluster richnesses, $\lambda_\textrm{MCMF}$, are estimated as a function of redshift from $0.01<z<2$, in steps of $\delta_z=0.005$. 
\item  These richnesses are computed as the sum of galaxy weights within a projected radius of $r_{500c}$ centered on the SZ candidate location where $r_{500c}$ is determined from the $\xi-M_{500c}$ relation (see Section \ref{sec:mass}) at the redshift of interest. 
\item Galaxy weights are computed for galaxies brighter than  $i \le m^{*}(z)$+1.25\footnote{Here $m^{*}$ corresponds to the apparent magnitude of L$^*$ galaxies, modeled as described in \citet{klein19}.}  and are based on (1) the consistency of galaxy colors and magnitudes with a redshift-dependent cluster population model and (2) a radial weight from the SZ center based on a normalized Navarro, Frenk and White (NFW) profile \citep{navarro97} with a scale radius of $R_s=r_\textrm{500c}/3$ \citep{hennig17}.  
Following e.g., \citet{gladders00}, a passive red-sequence population model is used to describe the color-magnitude relation of cluster galaxies. This model was empirically calibrated using ${\sim2,500}$ clusters with spectroscopic redshifts from the SPT-SZ cluster catalog \citep{bleem15b, bayliss16,khullar19}, the redMaPPer (RM) Y1 catalog \citep{rykoff16,mcclintock19},  and the MCXC cluster catalog \citep{piffaretti11}.
At low redshift, the 4000\AA \ break drives redshift determination, while at high redshift ($z>1.1$) where WISE data are used, the ``1.6 \um Stellar Bump'' feature (see Section \ref{subsec:spitzer}) provides discriminating power. 

\item A  correction is applied as needed to the richness at high redshift where the data are not complete to $m^{*}(z)$+1.25. 

\item Random sight lines are used to determine the probability of false associations as a function of redshift ($z_i$) and richness ($\lambda_i$), with this contamination fraction given by: 
\begin{equation}
f_\textrm{cont}(\lambda_i, z_i) = \frac{ \int_{\lambda_i}^{\infty}f_\textrm{rand}(\lambda, z_i)d\lambda}{ \int_{\lambda_i}^{\infty}f_\textrm{obs}(\lambda, z_i)d\lambda}.
\end{equation}
where  $f_\textrm{rand}$ is the richness distributions along random lines-of-sight and $f_\textrm{obs}$ is the richness distributions along candidate lines of sight. 

\item Up to three peaks per candidate are analyzed in richness-redshift space, with the peak associated with the lowest chance of being contamination assigned as the most likely counterpart to the SPT cluster candidate. 

\item Below a threshold of $f_\textrm{cont} = 0.2$ we denote candidates as ``confirmed''.  The expected net contamination in the confirmed sample is a combination of the intrinsic purity of the SZ sample (see Section \ref{subsec:expectedpurity}) and the optical/IR contamination. Under the assumption that the follow-up data is sufficiently deep to detect all real associations, the contamination of the optically confirmed sample is given by \citep[see also][]{klein23}: 
\begin{equation}\label{equation:opticalpurity}
\textrm{contamination}= f_\textrm{cont}^{\textrm{max}}*(1-p(\xi>\xi_\textrm{min}))
\end{equation} 
where $p(\xi>\xi_\textrm{min})$ is the purity of the complete SZ candidate sample at $\xi>\xi_\textrm{min}$, and we adopt $f_\textrm{cont}^{\textrm{max}}=0.2$.

\end{itemize}

\begin{figure*}
\begin{center}
\includegraphics[width=6in]{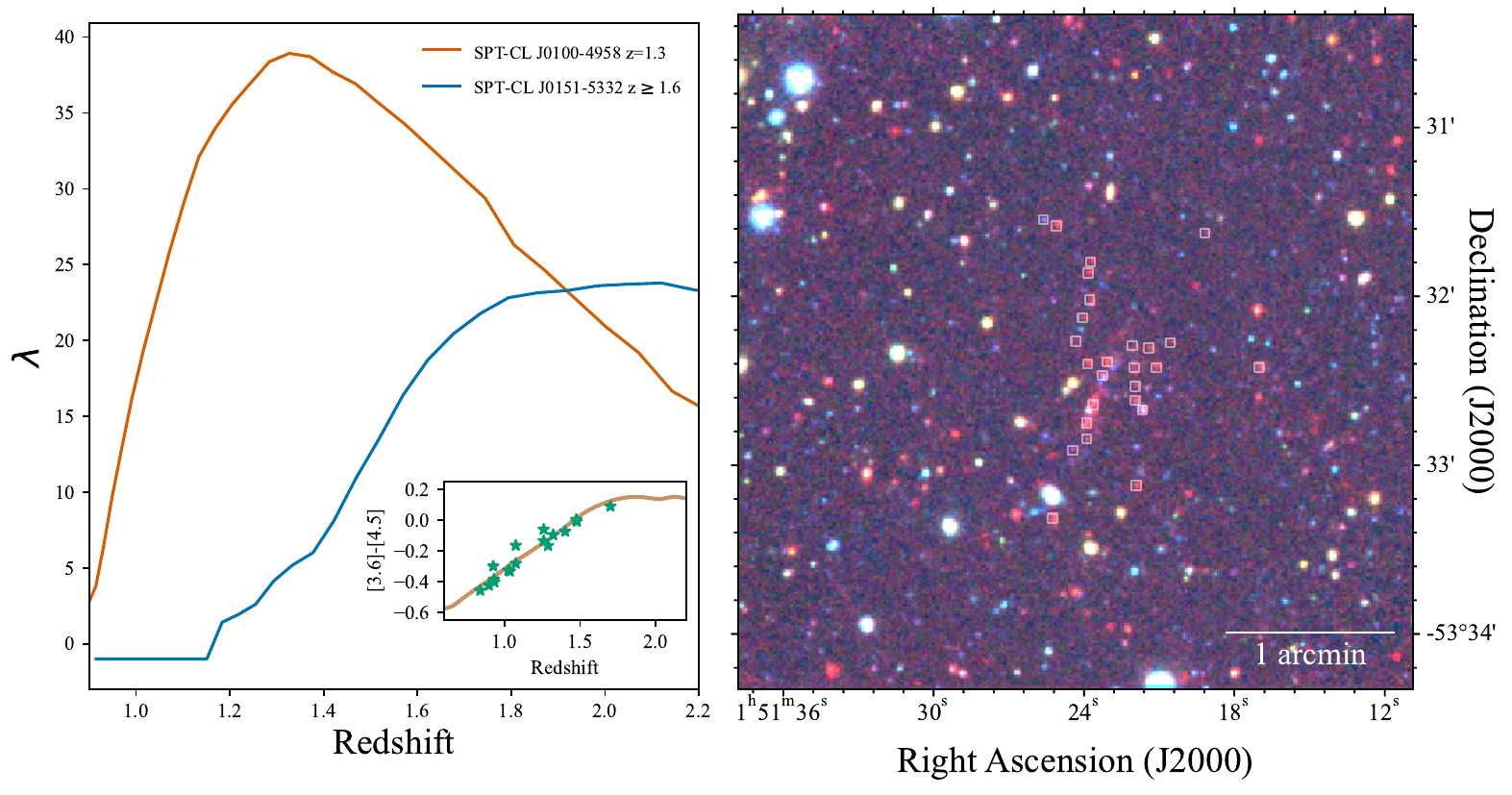}
\caption{\textit{(Left)} Example of \spitzer \ redshift determination on two high-$z$ SPTpol clusters. Plotted are IR richness ($\lambda$) versus redshift. In the inset we plot the median [3.6]$-$[4.5] color for SPT clusters with spectroscopic redshifts  (green stars) and  in brown we overplot our synthetic model calibrated using these systems. As can be seen, the \spitzer \ colors lose redshift discrimination power at $z>1.6$ and so we assign all systems with higher best-fit redshifts to $z=1.6$ and note this is a lower limit. \textit{(Right)} SPT-CL~J0151$-$5332 at $z\ge1.6$ with identified cluster members marked with white squares. The RGB image is constructed from \spitzer \ [3.6] and DES \textit{i}- and \textit{g}- band data.}
\label{fig:fitspitzer}
\end{center}
\end{figure*}

In Table \ref{tab:candidates} we provide MCMF summary statistics for the cluster candidates (except when confirmed by \spitzer, see below) including their redshifts and optical richnesses.

\subsection{Spitzer Confirmation}\label{subsec:spitzer}

For those cluster candidates with available higher-resolution and deeper \spitzer \ observations, we adopt a simpler, though related approach, to measuring the cluster redshifts, richnesses, and false associations. 
Given the small field-of-view of the targeted observations it is not possible to apply local background corrections. 

Clusters are confirmed in \spitzer \ data via the identification of excess galaxies at the cluster candidate locations as a function of [3.6]$-$[4.5] color via a variation of the ``1.6 \um Stellar Bump method'' previously employed in  e.g., \citet{papovich08, muzzin13b, gonzalez19}.  
As detailed in these previous works, after excluding low-redshift galaxies, there is a close mapping between the \spitzer \ color and redshift. 
We generate our model for this relation using the GALAXEV package \citep{bruzual03} assuming that the  cluster galaxy population was formed by a single starburst at $z=3$ with a Salpeter initial mass function \citep{salpeter55} and then followed the MILES \citep{vazdekis10} evolutionary tracks thereafter. 
As noted in e.g., \citet{sorba10}, the 1.6  \um feature 
is a robust feature immune to the details of the star formation history for all but the youngest stellar populations. 

We match sources selected from the \spitzer \ fields\footnote{Which we convert from Vega to AB magnitudes using the offsets of $m_\textrm{AB}=m_\textrm{Vega}+2.79 \ (3.26)$ for the [3.6]([4.5]) bands \citep{papovich16}.} to optically selected counterparts from the DES using a 1$\arcsec$ matching radius.  
Following \citet{muzzin13b,gonzalez19}, to reduce the number of low-$z$ interloper galaxies, we remove galaxies with $z-[3.6] <1.6$  and $i<21.3$ from our catalogs.  
We next run the single-color $\lambda$-richness estimator introduced in \citet{rykoff12} modified to use our 1.6 $\um$ bump redshift model in place of the red-sequence model. 
We adopt an intrinsic color-spread in the model at fixed redshift of $\sigma_\textrm{[3.6-4.5]}=0.07$ \citep{muzzin13b}. 
Background galaxy densities are estimated from either 25 blank-field pointings of unconfirmed cluster candidates from the SPT-SZ sample \citep{bleem15b} for the deep \spitzer \ observations or from the full SSDF field for the shallower observations. 
Following MCMF \citep{klein23}, we also modify the radial extent of the richness aperture to extend to what would be $r_{500c}$ for a cluster detected at significance $\xi$ at the redshift of interest.  
Richnesses were computed from $0.8<z<2$ with the candidate assigned the redshift and richness corresponding to the maximum $\lambda$ value. 
The [3.6]$-$[4.5] color loses redshift discrimination power at $z>1.6$ so systems with solutions at higher redshift are assigned $z=1.6$ and flagged in the catalog.  
An example of this fitting procedure for two different high-$z$ SPTpol clusters is shown in Figure \ref{fig:fitspitzer}.

While we do apply a color cut to remove low-redshift galaxies, faint galaxies below the DES detection limit can enter our catalog.  
As the 1.6 $\um$ bump redshift model above $z\sim0.7$ is degenerate with solutions at low-redshift \citep[see e.g., Figure 1 in][]{muzzin13b}, we also apply a secondary run estimating $\lambda$ versus redshift from $0.1<z<2$ and visually inspect all outputs, flagging and removing cases where low-redshift interlopers are biasing our results. 
In the vast majority of cases, these contaminants do indeed correspond to galaxies in the faint end of the luminosity function of 
rich low-redshift clusters confirmed by MCMF in DES optical data alone.  
A future improvement to our \spitzer \ confirmation work will incorporate additional optical information beyond our simple color cut. 

We also follow a similar convention to MCMF to estimate the contamination of the confirmed sample at fixed $\lambda$ but here restrict ourselves to a single wide redshift bin from $1<z<2$. 
To compute this contamination fraction we first select $\sim800$ random locations in the SSDF (with galaxies also matched to DES and the color cut to remove low-$z$ galaxies applied) that were screened to be $>5\arcmin$ from any SPTpol cluster candidate and not in highly masked regions of DES. 
We randomly assigned each location a $\xi$ value from the SPTpol catalog and then estimated $\lambda$ as for the real candidates to generate our random distribution. 
We denote fractional contamination determined using \spitzer \ data as $f_{\textrm{Scont}}.$ 
We adopt a somewhat more conservative  threshold ($f_{\textrm{Scont}}<0.1$ versus $f_{\textrm{cont}}<0.2$ from MCMF) for our \spitzer \ confirmation thresholds given the steepness of this contamination fraction with declining richness and our limited ability to sample blind \spitzer \ fields across the full 500d survey.

\subsection{Comparison Between WISE and Spitzer Observations}

\begin{figure}
\begin{center}
\includegraphics[width=2.75in]{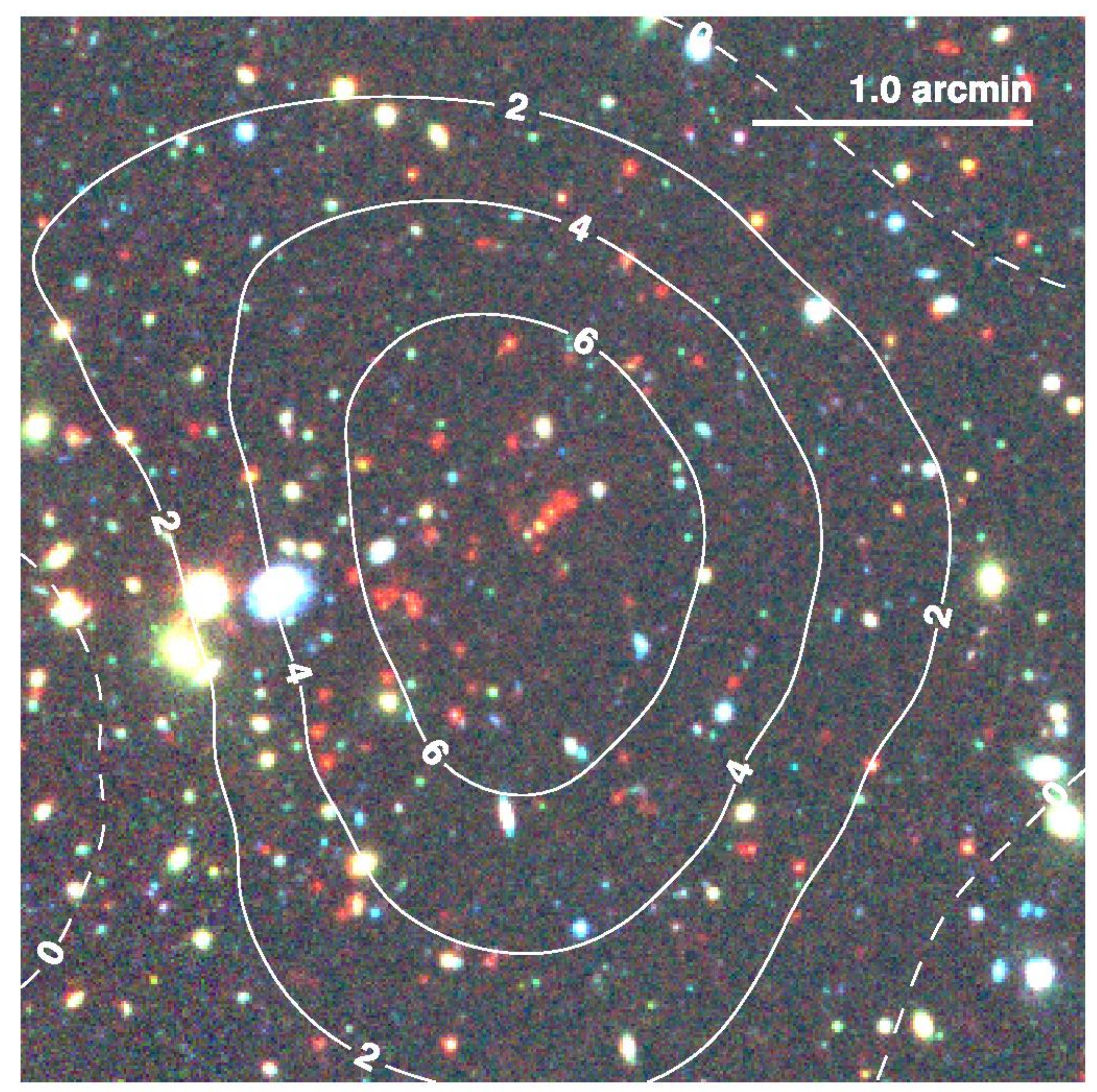}
\caption{Example of a system with multiple significant over-densities along the line-of-sight. SPT-CL~J2331$-$5737 at $\xi=7.6$ has potential counterparts located at (redshift, richness) $z=0.3, \lambda=65$ and $z=1.5, \lambda_\textrm{Spitzer}=25$. We adopt the higher-$z$ counterpart as our primary identification given its excellent alignment with the SPT position. While richer, the lower-$z$ counterpart is centered $\gtrsim70\arcsec$ away from SPT detection; the statistical uncertainty on the SPT position is 11\arcsec. The RGB image is constructed from \spitzer \ [3.6] and DES \textit{r}- and \textit{g}- band data.}
\label{fig:doublesystem}
\end{center}
\end{figure}

We can compare the performance of MCMF as applied to the DES-WISE dataset on high-$z$ clusters that were also observed in higher-resolution \spitzer /IRAC data. 
In total we have 108 candidates observed with \spitzer \ with estimated redshifts $z \ge0.85$ and \spitzer-derived contamination $f_{\textrm{Scont}}<0.1$. 
Of these systems, 78 are also confirmed via MCMF and have redshifts in decent agreement ($|\delta z| < 0.2$) with the \spitzer \ estimates. 
For the remaining 30 systems, differences in redshift/confirmation estimates between the MCMF and \spitzer \ analyses fall in two (expected) categories (1) detection in \spitzer \ data of higher-$z$ systems not well detected by MCMF in DES+WISE data, (2) identification of different galaxy over-densities along the line of sight, leading to different redshift estimates.
In the first scenario, 14 high-$z$ systems confirmed by \spitzer \ are not confirmed by MCMF ($f_{\textrm{cont}}>0.2$); 5 of these systems have MCMF redshifts approximately consistent with those obtained by \spitzer, but with richnesses up to a factor of 2$\times$ smaller, and 9 are solely detected in \spitzer \ data. 
The remaining 16 systems have significant differences in identified counterparts/redshifts between the algorithms. 
After further inspection, for 9 of these systems we selected the \spitzer-identified over-density as the primary counterpart as it is better centered on the SZ detection or a richer detection  and comparably centered (see Figure \ref{fig:doublesystem} and the middle panel of Figure \ref{fig:highzclusters} for two examples). For the remaining 7 systems the MCMF detection was selected as the primary counterpart. 
We flag systems with multiple significant over-densities along the line of sight in the cluster candidate table, Table \ref{tab:candidates}. 

In summary, we find generally good agreement between the MCMF/WISE and \spitzer \ analyses and that the WISE analysis provides an excellent addition to the analysis of DES data when deeper higher-resolution IR imaging is not available. 
The lower-resolution WISE data naturally has some limitations, and we find that the \spitzer \ data provide surer cluster confirmations (5 systems) or enables new detections not found in the WISE analysis (15 additional systems), impacting 20/108 clusters (16/51 at $z>1.1$) for which we had both datasets. 
In the near future data from the \textit{Euclid} mission \citep{laureijs11}
will enable us to reach beyond the limits of WISE confirmations for the full sample. 
These data, expected to allow detections of high-$z$ clusters to $z\sim2$ \citep{euclid19}, will enable confirmations of  additional high-$z$ systems from this SPTpol sample as well as future cluster samples from SPT-3G \citep{sobrin18}.

\section{Mass Estimation and Sample Purity}\label{sec:mass}
In this section, we detail how we connect our SPT observable, $\xi$, to mass. 
We also explain how we use realistic simulations of the SPTpol 500d field to explore the expected behavior of our observable-mass scaling relation and to estimate the purity of the SZ candidate list. 

\subsection{The $\zeta$-Mass Scaling Relation}

As in previous SPT cluster catalogs, we use an observable-mass scaling relation to relate our detection significances to mass. 
This is done through a two step process. First, as discussed in \citet{vanderlinde10}, to account for the bias owing to the maximization of the cluster detection algorithm over position and filter scales, we relate our observable $\xi$  to $\zeta$,  an ``unbiased"  detection significance via 
\begin{equation}
\label{eq:zetaxi}
P(\xi|\zeta) = \mathcal N(\sqrt{\zeta^2+3}, 1)
\end{equation}
for $\zeta > 2$ \citep[see also a formal derivation of this maximization correction in ][]{zubeldia21}. 

Next,  this unbiased significance is connected to mass via the relation: 
\begin{equation}
\langle\ln\zeta\rangle = \ln\bigl[A_\textrm{SZ} \left( \frac{\mass}{3 \times 10^{14} M_{\odot} h^{-1}} \right)^{B_\textrm{SZ}} \left(\frac{H(z)}{H(0.6)}\right)^{C_\textrm{SZ}}\bigr],
\label{eqn:zetam}
\end{equation}
and
\begin{equation}
P(\ln\zeta|M,z) = \mathcal N\left[\langle\ln\zeta\rangle(M,z), \sigma_{\ln\zeta}\right]
\end{equation}
where $A_\textrm{SZ}$ is the normalization, $B_\textrm{SZ}$ the slope, $C_\textrm{SZ}$ the redshift evolution, $\sigma_{\ln\zeta}$ the log-normal scatter on $\zeta$, and $H(z)$ is the Hubble parameter. 
As in \citet{reichardt13} and other SPT cluster works thereafter, we rescale the normalization to account for varying depth in the different SPT survey regions
\begin{equation}\label{eq:rescale}
A_\textrm{SZ} \rightarrow \gamma_\textrm{field}A_\textrm{SZ}.
\end{equation}

\subsection{Simulations of the SPTpol field}\label{subsec:sims}
We make use of simulations tailored to mimic the SPTpol observations to measure the rescaling factor, $ \gamma$, for the 500d field as well as to estimate the sample purity. 
Following \citet{bleem20, huang20}, we have constructed five realizations of the 500d field by combining realizations of the CMB \citep{keisler11}\footnote{We use \citet{keisler11} instead of more updated results from \citet{planck18-6} to maintain consistency with previous generations of SPT cluster simulations, but do note such a change would have negligible impact on the results presented here.}, cosmic infrared background \citep[CIB;][]{reichardt21}, discrete radio sources \citep{dezotti05} with spectral indices consistent with \citet{everett20},  thermal SZ, and  instrumental/residual atmospheric noise (the latter hereafter referred to as ``instrumental noise" for brevity).
The thermal SZ maps are created by applying the methods of \citet{flender16} to halo lightcones from the Outer Rim simulation \citep{heitmann19} and the instrumental noise maps 
are constructed from jackknife coadds of the SPT maps in which half of the observations were multiplied by -1 in order to produce maps with no sky signal.  
The SPT beam and transfer function are then applied to the simulated maps.
The emissive source subtraction on the real maps  (Section \ref{subsec:sourcesub}) reduces both the noise during cluster detection (by $\sim 3\%$ in the filtered maps) and the number of spurious cluster candidates. 
To mimic the effect of this subtraction in the simulations, we do not include radio sources brighter than the flux above which the source density of our radio source model matches the density of sources subtracted from the maps. 

Following \citet{reichardt13}, to measure $\gamma$, 
we first filter the maps with the same optimal filters used in the cluster identification and extract the $\zeta$ values at the known location and optimal filter scales of the simulated clusters. 
We then fit for the scaling relation parameters (Eq. \ref{eqn:zetam} above). 
We measure $\gamma=2.23$ for the SPTpol 500d field. 
This value implies the SPTpol 500d field depth should be roughly comparable to SPTpol 100d, and that clusters should be detected at $\sim$1.8$\times$ higher significance in SPTpol than SPT-SZ \citep[see e.g., Table 1 in ][]{dehaan16}.

We also tested the consistency of the $B_\textrm{SZ}$  and $C_\textrm{SZ}$ parameters measured in the SPTpol 500d simulations with those measured in simulated fields from the SPT-SZ \citep{bleem15b} and SPT-ECS \citep{bleem20} surveys. 
While we find $B_\textrm{SZ}$ to be consistent with previous values at better than the 1$\sigma$ level,  the best fit $C_\textrm{SZ}$ values have increased by $\Delta C_\textrm{SZ}=0.26$, corresponding to a 7$\sigma$ shift in the simulations. 

This change in $C_\textrm{SZ}$ arises from a combination of the reduced noise level and the larger contribution of the 95 GHz data to cluster detection in SPTpol compared to SPT-SZ. Decreasing the noise level increases sensitivity to higher-redshift clusters more than lower-redshift ones because of the fixed low-$\ell$ ``noise" contribution from the primary CMB. Meanwhile, the increased weight at 95 GHz improves sensitivity to low-redshift clusters. We have confirmed both of these trends in simulations: The values for  $C_\textrm{SZ,95 only}$ and  $C_\textrm{SZ,150 only}$ determined by running the cluster finder on single frequency maps are below and above the combined joint analysis value by 0.2, respectively, meanwhile excluding instrumental noise from simulations increases $C_\textrm{SZ}$. We surmise that, in the data, the effect of the noise reduction is larger, resulting in an increase in $C_\textrm{SZ}$.

\begin{figure}
\begin{center}
\includegraphics[width=3in]{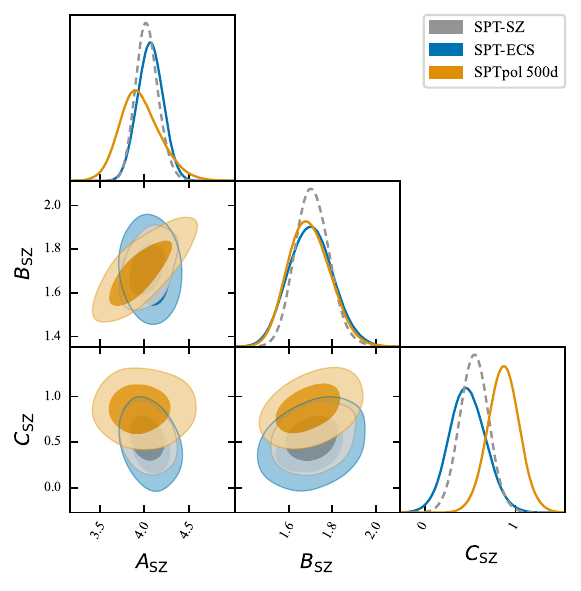}
\caption{Best fit mass-$\zeta$ scaling relation parameters (see Eq. \ref{eqn:zetam}) at our fiducial cosmology for the SPT-SZ, SPT-ECS, and SPTpol 500d cluster samples at $\xi>5$. There is good consistency between the normalization and slope  parameters ($A_\textrm{SZ}$ and $B_\textrm{SZ}$, respectively) in all three surveys; additionally the shift in the redshift evolution parameter ($C_\textrm{SZ}$) between SPTpol and the shallower surveys is captured in our simulations discussed in Section \ref{subsec:sims}. The smaller area surveyed in SPTpol 500d results in fewer massive clusters and increases the degeneracy between $A_\textrm{SZ}$ and $B_\textrm{SZ}$; this increases the uncertainty on the recovered $A_\textrm{SZ}$ parameter. } 
\label{fig:scalinggtc}
\end{center}
\end{figure}

\subsection{Scaling Relation Parameters}\label{subsec:parameters}
In recent works,
we have used scaling-relation parameters based on best-fit weighted averages from a Monte Carlo Markov chain  (MCMC) analysis of the abundance of clusters in the SPT-SZ dataset as a function of $\xi$ and redshift at our fiducial cosmology  \citep{bocquet19}. 
This analysis was conducted using the $\xi \ge 5$ sample for which we had complete optical/IR follow-up and assuming a fixed scatter of $\sigma_{\ln\zeta}=0.2$ whose value was motivated by comparisons with X-ray observables for a large sample of SPT clusters.
Motivated by observed changes in $C_\textrm{SZ}$ in our simulations, here we explore the validity of these scaling parameters for the SPTpol 500d sample using the data itself. 

In this test we replace the SPT-SZ sample above with the confirmed SPTpol 500d clusters at $\xi>5$ and run a new abundance analysis at fixed cosmology to estimate the SPTpol 500d mass-$\zeta$ scaling relation parameters.
The simulated normalization rescaling factors are applied for each survey field to make the normalizations directly comparable. 
We plot the results in  Figure \ref{fig:scalinggtc} along with the best fit parameters derived for SPT-SZ and SPT-ECS; the latter two results were derived using the $\xi>5$ confirmed samples from the respective surveys. 
We indeed observe a shift in the redshift evolution parameter of the scaling relation $C_\textrm{SZ}=0.87\pm0.17$ (compared to e.g., 0.64 $\pm$ 0.14 
in SPT-SZ) but find better than 0.5$\sigma$ \ consistency between the two surveys for the amplitude parameter $A_\textrm{SZ}=3.95\pm 0.23$ (4.08 $\pm 0.1$ in SPT-SZ) and mass slope $B_\textrm{SZ}=1.69\pm0.09$ (1.65$\pm$0.08 for SPT-SZ).  

Given the consistency in $A_\textrm{SZ}$ and $B_\textrm{SZ}$ for the 3 SPT surveys of significantly different depths, and the increased degeneracy between these parameters in the SPTpol 500d field given the smaller number of massive clusters observed owing to its smaller survey volume, 
we continue to use the best fit values from SPT-SZ for these parameters. 
Given the significant shift in the $C_\textrm{SZ}$, we adopt the newly derived value from SPTpol 500d data  when reporting our cluster masses for the SPTpol 500d sample.

\subsection{Expected Purity of the SPTpol 500d Cluster Sample}\label{subsec:expectedpurity}
In this work, we apply a new method to estimate the cluster sample purity that significantly improves the accuracy at lower detection significances.  
In previous SPT works, the number of false candidates as a function of significance was estimated by running the cluster detection algorithm  on sky simulations with no tSZ signal. This is sufficient for characterizing the properties of high-significance samples drawn from maps in which significant instrumental and residual atmospheric noise is present (such as the SPT-SZ sample presented in \citealt{bleem15b} and used in the cosmological analyses of \citealt[][]{dehaan16,bocquet19}). 
However, it is known that astrophysical foregrounds and the tSZ itself are not necessarily Gaussian \citep[see e.g., recent measurements in][]{crawford14,coulton18}. 
The impact of this non-Gaussianity on the expected number of false candidates becomes more pronounced as cluster samples are produced from lower-noise data and to lower detection significance.

\begin{figure}
\begin{center}
\includegraphics[width=3.0in]{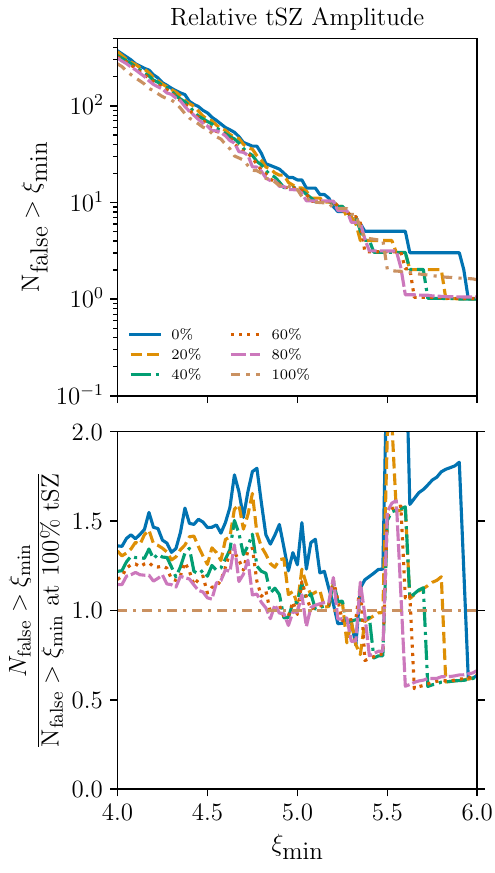}
\caption{Variations in expected number of false detections based on differences in assumed tSZ amplitude for a 500-square-degree SPTpol-like field. In the \textit{(top)} panel we show the expected number of false detections at $\xi>\xi_{\textrm{min}}$ and in the \textit{(bottom)} panel the ratio in the number of spurious detections compared to our reference assumptions. We vary the relative amplitude of the tSZ signal in the sky simulations, where ``100\%" corresponds to the maps having tSZ power which matches SPT power spectrum data \citep{reichardt21} at $\ell=3000$.  Estimating the expected number of false detections through tSZ-free simulations can overestimate their counts by factors of 1.4 at low-$\xi$.}
\label{fig:false_variations}
\end{center}
\end{figure}

To illustrate this effect, we measure the number of false candidates in a simulated 500 \sqdeg \ region varying the amplitude of the tSZ signal. 
The process by which we identify false candidates is discussed below. 
We plot the results of these simulations in Figure \ref{fig:false_variations}.
As can be seen, the relative amplitude of the tSZ to the total noise can have a significant impact on the expected number of false detections, especially at lower $\xi$ values; using  tSZ-free simulations this quantity can be overestimated by factors of $\sim1.4$ at $\xi_{\textrm{min}}=4$.  
This tension with estimations from tSZ-free simulations at low $\xi$ is also seen empirically using optical follow-up observations as discussed in \citet{klein23}.

In light of this effect, we update our method of prediction for the expected number of false candidates (and correspondingly sample purity) to use simulations that are statistically as close to the data as possible.
We run the cluster finder on maps with tSZ included and make use of the fact that our tSZ maps are constructed with perfect correspondence to massive halos in the OuterRim Simulation. This provides us with a complete listing of all real possible cluster detections. 
We then estimate the purity of our cluster sample at  $\xi>\xi_\textrm{min}$ as follows:

\begin{itemize}
\item For each of 5 independent simulated realizations of the SPTpol 500d field (2500 square-degrees in total) we run the cluster detection algorithm. Each realization is drawn from a non-overlapping region of the simulated Compton-$y$ map and has different CMB, foreground, and noise realizations. Starting with the highest-significance cluster detection, candidates are matched to the most massive unmatched simulation halo within 2 arcmin. Matched halos are prevented from being matched with lower-significance candidates. 
\item A similar matching is done for 5000 random sightlines within each field footprint which determines the probability of randomly associating with a halo of a  given mass within the association radius, $f_\textrm{R}(M)$. Because of the strong dependence of number density on halo mass owing to the steepness of the halo mass function, it is important to include mass information when quantifying random associations.
\item Next, using the observed number of candidates at $\xi>\xi_\textrm{min}$ as well as this random association probability as a function of halo mass, we can estimate the number of false candidates above $\xi_\textrm{min}$.  Starting at the highest mass (with lowest chance of random association), we calculate the fraction of true associations to total candidates, $p(M,\xi_\textrm{min})$, given $N_\textrm{obs}$ associations for $N_\textrm{cand}$ candidates at significance $>\xi_\textrm{min}$ as 
\begin{equation*}
p(M,\xi_\textrm{min}) = \frac{N_\textrm{obs} - N_\textrm{cand}f_\textrm{R}(M)}{N_\textrm{cand}(1-f_\textrm{R}(M))}
\end{equation*}
The number of false associations, $N_\textrm{FA}$, at $\xi>\xi_\textrm{min}$ and mass is then simply 
\begin{equation*}
N_\textrm{FA}(M,\xi_\textrm{min}) = 
N_\textrm{obs} - p(M,\xi_\textrm{min})N_\textrm{cand}
\end{equation*}
We integrate with decreasing mass, reducing the number of candidates available for random associations by the true associations calculated in previous steps, to compute the total number of false candidates. This value is given by the sum of cluster candidates unassociated with simulated halos and the number candidates falsely associated with such halos. 
\item We then estimate the purity of the sample in the real survey by dividing this number of false candidates by the observed number of candidates in the SPTpol sample. 
\end{itemize} 

\begin{figure}[h]
\begin{center}
\includegraphics[width=3.5in]{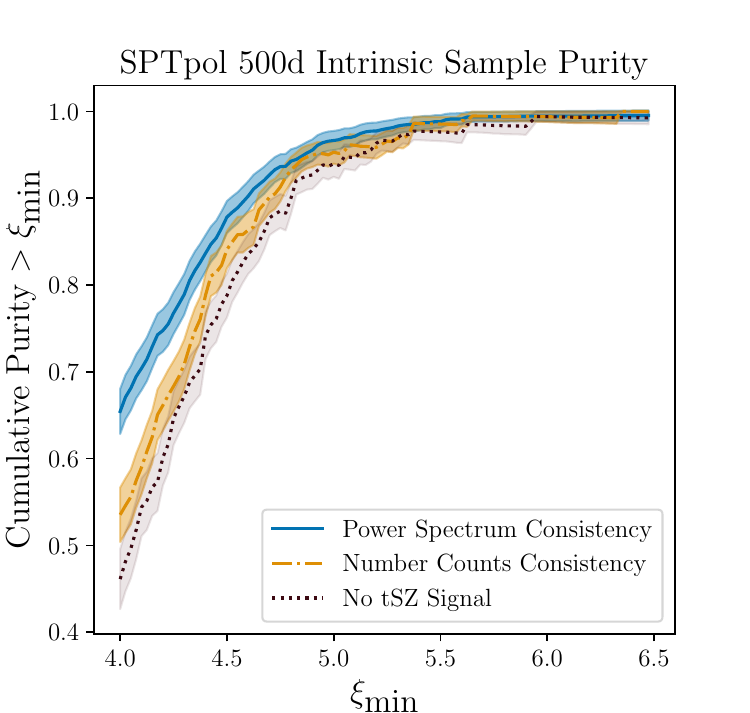}
\caption{Expected intrinsic purity of SPTpol 500d candidate list (without considering optical/IR confirmation) under different modeling assumptions for the tSZ contribution to simulated survey maps. Overall our purity for the SPTpol sample agree within 1 $\sigma$ at $\xi>4.5$  for our two bounding assumptions of either (1) matching the tSZ amplitude at $\ell=3000$ to recent power spectrum constraints \citep{reichardt21} or (2) matching the identified candidate density in the 500-sq-degree footprint to that of the observed SPTpol sample. Estimating the number of false detections (and hence purity) of the sample using tSZ-free simulations significantly underestimates the sample purity at lower $\xi$ values.  }
\label{fig:purity_fig}
\end{center}
\end{figure}

We have estimated this purity under two different assumptions that we plot in Figure \ref{fig:purity_fig}. 
The first  uses simulations in which we have normalized the tSZ power spectrum at $\ell=3000$ to match previous SPT results \citep{george15, reichardt21} and we adopt this as our baseline model. 
This tSZ model leads to approximately twice as many clusters at $\xi \ge 4$ in the simulated SPTpol field as compared to observation (likely owing to some combination of different cosmological parameters between the simulations and real data,  differences from reality of the assumed tSZ profile as a function of mass and redshift, and lack of simulated correlated emission from cluster members with the tSZ signals, though we find no strong evidence for the latter in our data---see Section \ref{sec:biases}). 
As an alternative, we estimate the purity using simulations where we have scaled the tSZ amplitude so that the candidate count matches the observed SPTpol results. 
These two models agree with each other within 1 $\sigma$ at $\xi>4.5$.
For reference we also plot the ``tSZ-free" simulation case, which shows lower purity estimates at low $\xi$ than the other two models. 

Using Eq. \ref{equation:opticalpurity}, our baseline SZ purity model,  and the subset of cluster candidates with contamination {$<0.2$}, we estimate an overall purity of our confirmed cluster sample of 93\%. 
We can omit use of the simulations, and estimate a purity such that Eq.  \ref{equation:opticalpurity} is satisfied and we observe consistency between the number of observed unassociated detections and predictions \citep[as was done in][]{klein23}. This estimate raises the expected purity value slightly to 95\% which corresponds to a $\sim2.5\sigma$ difference between the two estimates of the purity of the full SZ candidate sample $\xi>4$. 
Given the good agreement between the estimates for the confirmed portion,  we quote $\sim94\%$ as the expected purity of our confirmed sample. 
To further refine our purity estimates and cluster selection modeling we are in the process of improving our simulations of cluster gas properties \citep{keruzore23} and correlated sources of mm-wavelength contamination. This work will be important to develop in parallel to keep pace with the upcoming cluster samples from e.g., SPT-3G \citep{sobrin18}, and Simons Observatory \citep{ade19}. 

\section{Systematic Explorations}\label{sec:biases}
We conduct several tests to search for potential biases in the SZ signal we use to identify SPTpol clusters and estimate their masses. 
Such biases may arise from correlated emission from cluster galaxies, namely synchrotron from radio sources or emission from dust associated with star formation in cluster members. 
Based on multi-wavelength studies of the cluster galaxy population \citep[see e.g.,][]{gralla11,alberts16,melin18}, synchrotron contamination is expected to arise from bright discrete sources such as active galactic nuclei in cluster central galaxies, while contamination correlated with star formation is predominately sourced through the integrated emission from a large number of cluster members.  
Here we use external radio imaging data as well as internal comparisons of data from the two SPTpol frequencies to search for and estimate the level of bias from these sources in our sample. 
We note that as this study is being conducted using an SZ-selected sample, the impact of very strong contamination (such as from a powerful radio source) that completely fills in the SZ decrement will not be accounted for in this test. 
Based on previous studies of optical-, IR-, and X-ray-selected systems \citep[e.g.,][]{lin09, gralla11,gralla14,gupta17,mo20,dicker21}, such extreme sources are expected to be rare and not have a large impact on the completeness of SZ surveys at redshifts $z>0.25$. 

\subsection{Radio Source Check}\label{sec:radio_source}

To assess potential contamination from radio sources below the detection threshold
of SPTpol, we follow a procedure very similar to that used in \citet{bleem20}. In this
work, we use publicly available thumbnail maps from
the Sydney University Molonglo Sky Survey (SUMSS, \citealt{mauch03}) at 843\,MHz, 
while \citet{bleem20} used the 1.4~GHz 
National Radio Astronomy Observatory (NRAO) Very Large Array (VLA) Sky Survey
(NVSS, \citealt{condon98}) which unfortunately does not extend as far south as the SPTpol field. 
Otherwise the process is identical; we summarize
the analysis briefly here and refer the reader to \citet{bleem20} for more details.

First, we download all SUMSS postage-stamp maps that overlap with the SPTpol
500d field and reproject them onto the same pixel grid as the SPTpol maps. We
make beam- and transfer-function-matched SUMSS maps for each of the SPTpol observing
frequencies and scale the intensity of the maps assuming a single spectral index
of $-0.7$ \citep{coble03}, and we convert the result to CMB fluctuation temperature. We produce 
maps of contamination to the cluster-finding by combining the single-frequency maps
with the same weights as used in the cluster-finding and then filtering the result with 
each of the 12 matched filters. We estimate the contamination to $\xi$ for each 
cluster by taking the value of the appropriate contamination map at the cluster center 
and dividing by the noise in the actual SPTpol combined, cluster-filtered map at that 
location. Because of artifacts in the SUMSS map around the bright radio source 
PKS~2356-61, we are not able to perform this calculation within approximately 
1.5 degrees of that source. Eight of the 689 candidates in the SPTpol 500d catalog 
lie in this area. 

The median contamination calculated in this way is $\Delta \xi_\mathrm{med} = 0.032$ (where $\Delta \xi$ is defined such that $\xi_\mathrm{observed} = \xi_\mathrm{true} - \Delta \xi)$, or 0.8\% of the $\xi=4$ threshold value for inclusion in the catalog. 
Of the 681 candidates in the catalog for which we are able to perform this test, 
47 ($\sim$7\%) have a predicted contamination of greater than $\Delta \xi = 1$, and 
18 ($\sim$2.5\%) have a predicted contamination of greater than $\Delta \xi = 2$.
Of the 18 candidates with a predicted contamination greater than $\Delta \xi = 2$, three
are low-redshift ($z < 0.25$) systems, and two have $\xi < 4.25$, leaving only thirteen
such candidates that would be included in a cosmological analysis.
We flag candidates with predicated contamination $\Delta \xi > 2$ in the candidate Table \ref{tab:candidates}. 

\begin{figure}
\begin{center}
\vspace{0.18in}
\includegraphics[width=3in]{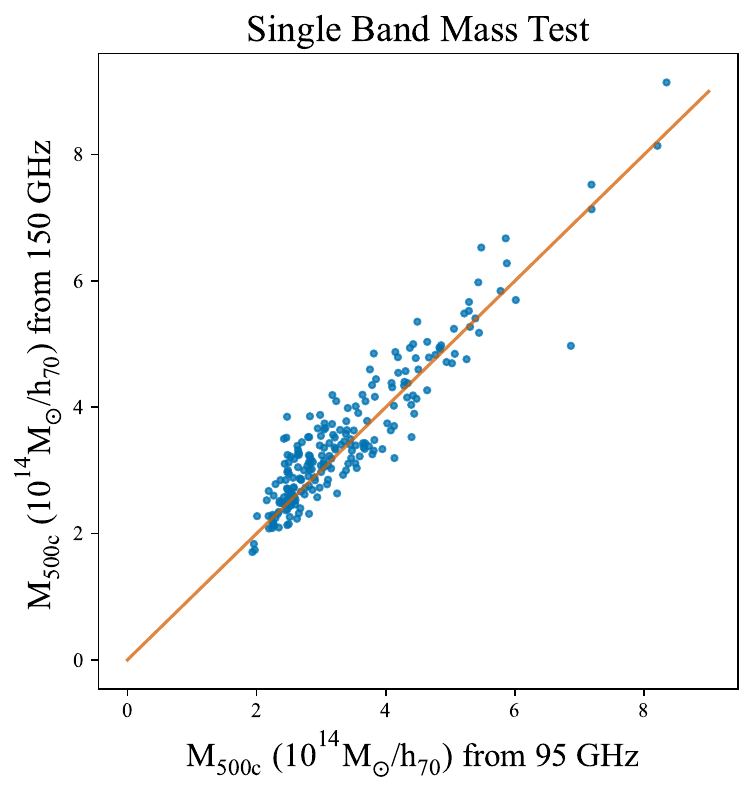}
\caption{Masses estimated from 95 and 150 GHz data alone for 250 clusters individually detected in each band. Overall we find excellent agreement between the two, with the largest outlier,  SPT-CL~J2332-5358 (5.4$\sigma$, at a ratio of M$_{\textrm{95 GHz}}$/M$_{\textrm{150 GHz}}=7/5$) having previously been identified as a cluster-scale strong lens of a distant star forming galaxy whose emission results in significantly reduced tSZ signal  in 150 GHz data.}
\label{fig:mass95150}
\end{center}
\end{figure}

\subsection{95/150 GHz Internal Consistency Test}\label{subsec:internal}

We can use cluster samples selected independently from the SPTpol 95 and 150 GHz data to provide an independent test of predictions from our simulations as well as to look for signatures of potential contamination to the SZ signal.
This test is based on the premise that the two key expected sources of correlated contaminating emission---synchrotron and  thermal dust radiation from cluster member galaxies---should impact the recovered SZ differently at each frequency given their differing spectral energy distributions.  
Here we compare end-to-end predictions of the masses estimated from each frequency as cluster mass is what we are most concerned about for cosmological analyses. 

Following the cluster identification procedure of Section \ref{sec:clusterfinding} but now run on the single band maps, we detect 540 candidates (514 candidates) at $\xi>4$ in the 95 (150 GHz) data with 250 common detections representing 221/\nconfirm \ confirmed systems with redshifts. 
Repeating the simulations above to standardize the extracted $\xi$ values we expect the amplitude normalization factors, $\gamma$, to be 0.85 (0.78) at 95 (150) GHz relative to the full depth field, with changes in $C_\textrm{SZ}$ between the bands as discussed above.

In Figure \ref{fig:mass95150} we plot the masses estimated from each individual frequency for the matched clusters. 
We find the median ratio in the masses estimated from 95 and 150 GHz alone to be  $0.968\pm 0.005$. 
We attribute this offset from unity to a small misestimation of the foreground levels in the simulations, due to uncertainty in the amplitude of the uncorrelated CIB at 95 GHz \citep{reichardt21}, as well as our simplified modeling of the applied source cleaning, which might affect the bands differently due to the source populations in each map.  For example, if we split the sample into halves based on the estimated level of radio contamination from the SUMSS data, the 95/150 GHz mass ratio is only 1\% different between the two halves.  Regardless, as the amplitude and other parameters of our observable-mass scaling relation will be determined via weak lensing \citep[e.g., ][]{dietrich19,schrabback18,schrabback21}, small shifts of this order in our simulation calibration have negligible impact.

We next check for a redshift evolution in this relation. Splitting this joint sample at its median redshift of $z=0.66$, we find the median ratio of the masses to be consistent, $0.97^{+0.01}_{-0.02}$ for the lower redshift half and $0.97\pm0.01$ for the higher redshift half. Restricting ourselves to the highest redshift clusters in the common sample at $z>1$ (35 systems) we may see some evolution in the CIB contribution to the 150 GHz band, with the median ratio of 95/150 GHz masses shifting to $1.04^{+0.02}_{-0.03}$, indicating that the 150 GHz signal may be being partially filled in.  
This change would result in a $\sim3.5\%$ change in the estimated masses of these highest redshift systems in our multi-band mass estimates, and is much smaller than our current best fit uncertainty in the $C_\textrm{SZ}$ relation calibrated with weak lensing \citep{schrabback21,zohren22} or constrained by our joint lensing+cosmological analyses \citep{bocquet19}.  

To check for outliers in the mass comparison, we difference the masses determined by data from each frequency and divide by the mass uncertainty determined by the statistical uncertainty only (as the intrinsic Compton-$y$ mass  scatter should be the same for a given cluster measured at two different frequencies). 
To account for correlated scatter in the noise at each frequency from common foreground/atmospheric residuals we normalize this distribution to be a Gaussian of unit width by reducing the statistical uncertainty by 0.57, an empirical factor measured by taking a robust measure  of the standard deviation of the distribution.\footnote{This robust standard deviation is computed using the median absolute deviation as an initial estimate, and then using Tukey's biweight to weight points, see discussion in e.g.,  \citet{beers90}.} 
We find three clusters where the 95 and 150 GHz mass estimates differ at $>3\sigma$.
The largest discrepancy is  SPT-CL~J2332-5358 (5.4$\sigma$) which was previously discussed in \citet{vanderlinde10,Andersson11} as a cluster that is lensing a distant star-forming galaxy. The emission from this galaxy significantly reduces the SZ signal measured at 150 GHz which lowers its 150 GHz mass estimate. 
This cluster is noticeable as the biggest outlier in Figure \ref{fig:mass95150}.

Conversely, there are two systems (SPT-CL~J2337-5942, SPT-CL~J0154-5619) both with 95 GHz mass estimates less than the 150 GHz mass estimates by $\sim$3.1$\sigma$.
These differences might be caused by co-located radio sources.  
Indeed for SPT-CL~J2337-5942 there are 3 low-brightness SUMSS sources (7-10 mJy) within 2$\arcmin$. 
However, for SPT-CL~J0154-5619, there is no SUMSS source within 2$\arcmin$ of the SPT location. 
  Given the sample size of 221 clusters, one would expect two systems as $\ge 3.1\sigma$ outliers $\sim7\%$ of the time.

If we look instead at clusters detected significantly in one channel  but not the other ($>7\sigma$, or 3$\sigma$ above our search threshold), we find only one system, SPT-CL~J2240$-$6117 at $z=0.95$, detected in 95 GHz data at $\xi_{\textrm{95}}=8$. Using forced-photometry on the 150 GHz detection map, we find the cluster detected at $\xi_{\textrm{150}}=2.2$. 
Examination of a preliminary version of the first SPT-3G point source catalog (Archipley et al., in prep), reveals a dusty source detected at 220 GHz at 15 mJy  
within 0$\farcm{2}$ \ of the 95 GHz centroid, making this system potentially a cluster lens similar to  SPT-CL~J2332-5358 discussed above. 
The high sensitivity of the SPT-3G at 220 GHz, in combination with significantly deeper data 95 and 150 GHz, will offer more opportunities to identify instances of high-$z$ dusty sources lensed by massive clusters. 

\begin{figure*}[t!]
\begin{center}
\includegraphics[width=7in]{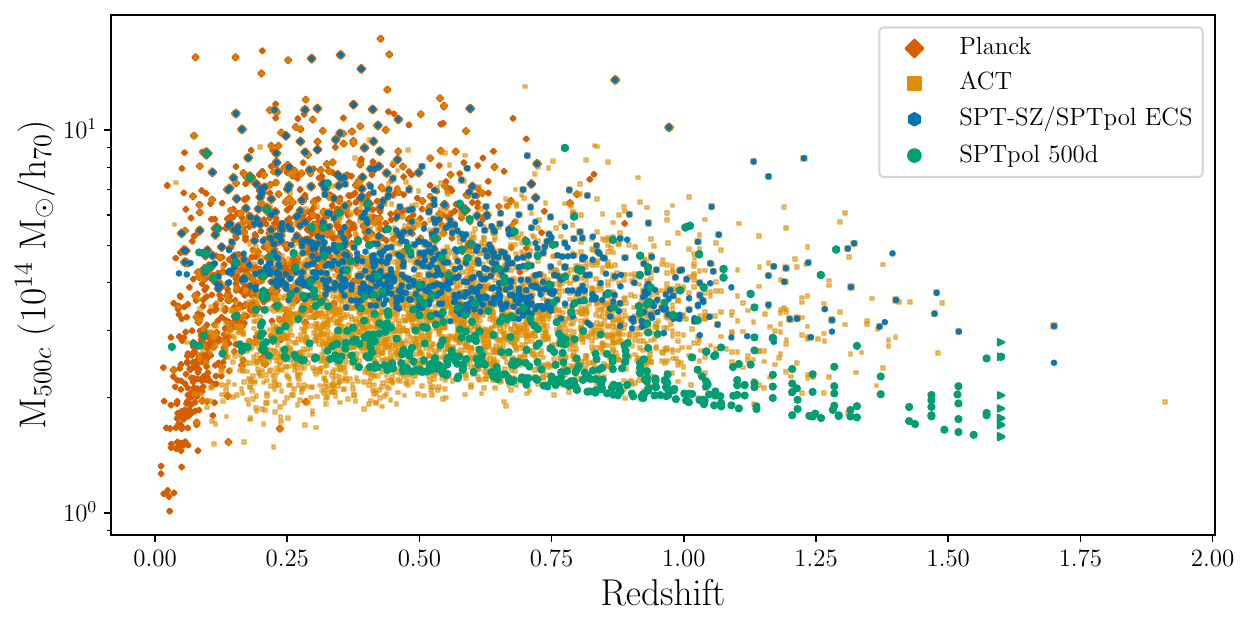}
\caption{The mass-redshift distribution of the SPTpol 500d cluster sample. Plotted for comparison are wide-field SZ cluster samples from \planck \ \citep{planck15-27}, ACT \citep{hilton21}, and SPT-SZ/SPT-ECS \citep{bleem15b,bleem20}. The SPTpol sample consists of \nconfirm \  clusters, with 21\% of the sample at $z>1$. As discussed in Section \ref{subsec:spitzer},  redshifts for clusters confirmed in \spitzer \  and WISE data at $z\ge1.6$ are considered lower limits. These systems are plotted as right-facing triangles in the plot.}
\label{fig:massplot}
\end{center}
\end{figure*}

\section{The 500-square-degree SPTpol Survey SZ Cluster Catalog}\label{sec:catalog}

The SPTpol 500d cluster catalog consists of 689 SZ candidates detected at $\xi>4$. 
Using  optical and infrared observations we have confirmed \nconfirm \ of these candidates as galaxy clusters with an expected contamination of our confirmed sample of less than $\sim6\%$. 
The redshifts  of the confirmed sample are in the  range  $0.03  < z \lesssim 1.6$ and the masses are in the range $1.5  \times 10^{14} < M_{500c} < 9 \times 10^{14} $\msun.
The sample has a median redshift of 0.7, median mass of $2.5 \times 10^{14} $\msun, and a spatial density of 1.18 confirmed clusters/\sqdeg. 
We provide the complete cluster candidate list as well as redshifts, estimated masses, and select optical properties for confirmed clusters in Table \ref{tab:candidates}.

\begin{figure*}
\begin{center}
\vspace{0.35in}
\includegraphics[width=7in]{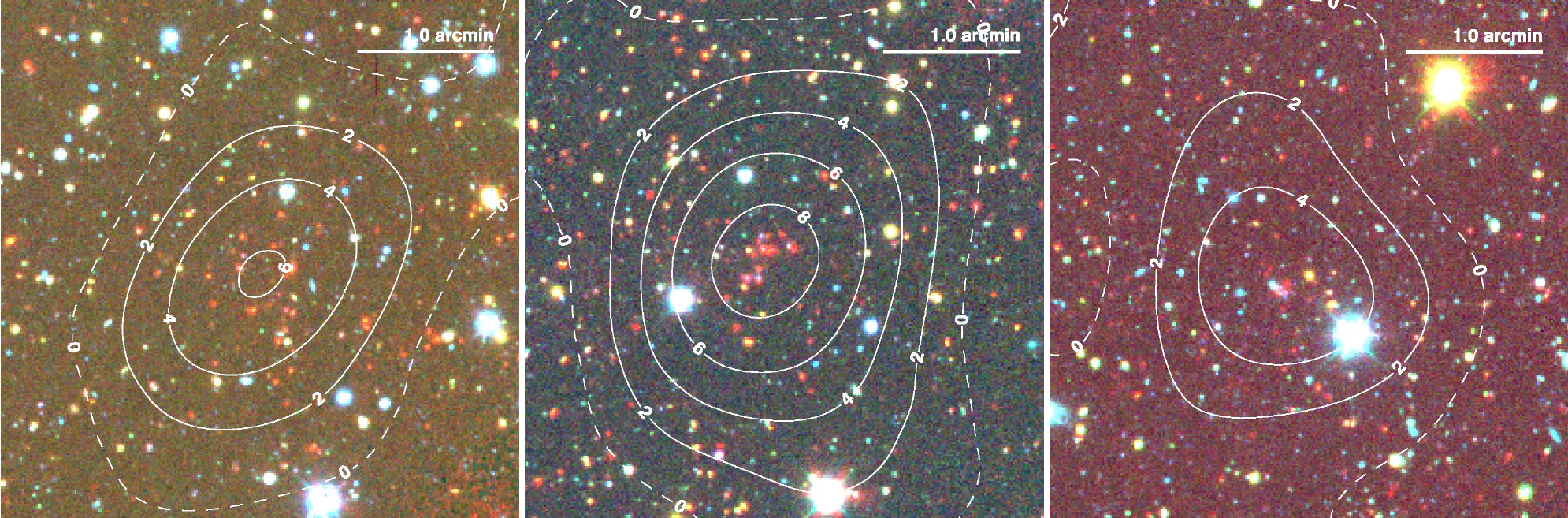}
\caption{Three of the high-$z$ clusters discovered in the SPTpol 500d survey. From left to right, SPT-CL~J0128-5222 at $\xi=6$ and $z=1.4\pm 0.07$, SPT-CL~J0122-5801 at $\xi=9$ and $z\gtrsim1.6$, and SPT-CL~J0116-5039 at $\xi=6$ and $z\gtrsim1.6$. The SZ detection contours are overlaid on RGB images from \spitzer \ [3.6] and DES \textit{i}- and \textit{g}- band data. Star formation---as traced by bluer emission correlated with the galaxy overdensities---is  prominently visible in some of these high-$z$ cluster galaxies, highlighting the ability of this sample to probe clusters in the high-$z$ transitional era between active star formation and passive galaxy evolution.  SPT-CL~J0122-5801 is one of the systems tagged as having multiple overdensities along the line-of-sight, with an additional foreground system of $\lambda=16, z=0.6$, and $f_\textrm{cont}=0.05$. The deep high-resolution \spitzer \ data allow us to detect the more significant distant background cluster. }
\label{fig:highzclusters}
\end{center}
\end{figure*}

In Figure \ref{fig:massplot} we plot the mass-redshift distribution of the SPTpol 500d sample as compared to several other SZ-selected cluster samples. 
As seen in this plot, the high sensitivity of SPTpol and our extensive follow-up efforts have enabled us to confirm a large number of low-mass and high-redshift clusters; 114 of the SZ clusters are at $z>1$ (21\% of our total sample).  
This is a much higher fraction than the SPT-SZ \citep[42/516 or 8\%,][]{bleem15b} and ACT samples \citep[222/4195, 5\%,][]{hilton21}. 
We further compare select properties of the SPTpol sample to these other samples in Section \ref{sec:comparison}. 
The redshift and mass reach of the SPTpol sample will enable a number of exciting studies of both cosmology \citep[e.g.,][]{bocquet23} and astrophysics through studies of the mass and redshift evolution of clusters. 
In Figure \ref{fig:highzclusters} we highlight several of the new high-$z$ clusters reported in this work. 

\begin{figure}
\begin{center}
\includegraphics[width=3in]{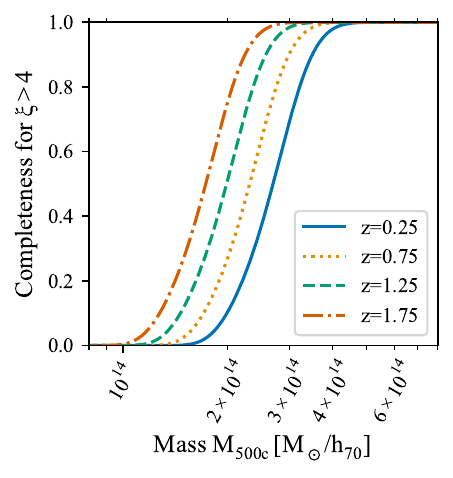}
\caption{Expected completeness of the $\xi>4$ SPTpol sample as a function of $M_{500c}$ at several redshifts. The sample is expected to be $>90\%$ complete at $M_{500c} >3.5\times 10^{14}$ \msun \ at $z>0.25$. At lower redshifts the atmospheric filtering removes larger angular-scale cluster signals leading the sample to be increasingly incomplete at $z<0.25$.}
\label{fig:selection_function}
\end{center}
\end{figure}

In Figure \ref{fig:selection_function} we plot the estimated selection function of the sample at $z>0.25$. 
This selection function is empirically derived via propagating the $\xi \ge 4$ sample cut into mass- and redshift-space using the $\xi$-mass relation discussed in Section \ref{sec:mass}. The SPTpol sample is highly \mbox{($>90\%$)} complete at masses $M_{500c} > 3.5 \times 10^{14} $\msun  \ and $z>0.25$.
The SPTpol survey, like previous SPT works, has an increasing sensitivity to lower-mass clusters as a function of increasing redshift. 
As described in \citet{huang20}, this trend arises mainly from two effects: 
(1) residual fluctuations from the CMB and atmosphere increase the noise in the maps at larger angular scales (2) self-similar evolution of clusters leads to hotter clusters at fixed mass at higher redshifts and hence makes higher-$z$ systems easier to detect.

\subsection{Comparison to Other Cluster Surveys}\label{sec:comparison}
In this section, we compare the properties of the SPTpol 500d cluster sample to SZ and optical samples in the same survey region. 
For the SZ samples we focus on catalogs from ACT and SPT\footnote{While there is overlap of 22 systems with  \planck, the typically higher redshift and lower mass SPTpol 500d sample adds no significant new information to the SPT-\planck \ comparisons previously undertaken in \citep{bleem20} for the SPT-SZ and SPT-ECS samples.} and check the consistency of mass and redshift estimates for the clusters in common between the samples. 
For our comparisons to the optically selected systems from DES and WISE data we check whether the SZ systems were detected and compare estimated redshifts. 
Detailed comparisons that more fully leverage the constraining power offered by the lower-mass SPTpol systems, such as exploring the $\lambda-\mass$ relation and contamination of the optical samples \citep[see, e.g.,][]{bleem20,grandis21}, are reserved for future work.

\begin{figure*}
\begin{center}
\includegraphics[width=6in]{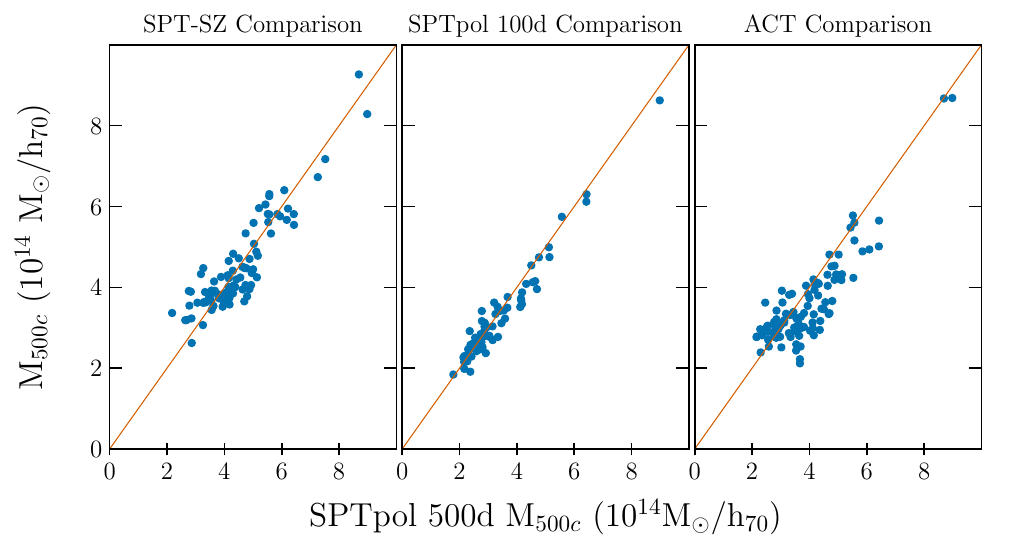}
\caption{Comparisons of mass estimates for clusters from SPTpol 500d that are in common with systems in SPT-SZ \citep{bleem15b}, SPTpol 100d \citep{huang20}, and ACT \citep{hilton21}. Over-plotted is a line representing a one-to-one relationship. As discussed in Section \ref{sec:comparison}, we find good agreement between the 3 SPT surveys, with the median ratio of masses of  SPTpol 500d/SPT-SZ to be $1.02^{+0.02}_{-0.04}$ and SPTpol 500d/SPTpol 100d to be $1.02 \pm 0.02$. We do note a small ($2.3\sigma$) difference in the normalization of the ACT and SPT masses, finding the ratio of masses from SPTpol 500d/ACT =$1.07 \pm 0.03$ \citep[differing from the consistency shown in][]{hilton21}, but a stronger test of any cross experiment mass calibration differences will come when comparing weak-lensing calibrated samples in future works. }
\label{fig:mass_comparison}
\end{center}
\end{figure*}

\subsubsection{SPT-SZ}\label{sec:sptsz_comparison}
There are 118 SPT-SZ cluster candidates detected at $\xi>4.5$ that fall within the SPTpol 500d footprint  and 113 of these candidates in unmasked regions (with masking owing to nearby bright point sources as discussed in Section \ref{subsec:processing}).\footnote{We note the point source mask between SPT-SZ and SPTpol differs slightly owing to survey noise and source variability as each was constructed using sources brighter than {$\sim$6} mJy as measured in the maps used for cluster identification.}
Using an association radius of $3\arcmin$ to identify matching detections, we find that 97 of these candidates have matches in the SPTpol catalog, with median ratio of the significance of cluster detection between SPTpol/SPT-SZ being 1.7, and a median spatial separation of 0$\farcm{26}$. 
As discussed in \citet{klein23}, for 5 of these matched systems, the improved DES and WISE follow-up data allow us to confirm them and provide redshifts. 
Of the 16 unmatched SPT-SZ candidates ranging in $4.5<\xi<5.2$ (median $\xi=4.7$), only one---SPT-CL~J2232-6151, $\xi=5.04, z=0.79$---was reported confirmed by follow-up optical and infrared observations in \citet{bleem15b}.
This is consistent with the estimate for the purity of the optically confirmed sample, using the framework developed in this paper. 
There are  597 cluster candidates at $\xi>4$ in SPTpol that are not found in SPT-SZ, with 448  (385) of these having optical contamination values less than 0.2 (0.05). 
Of these systems, 123 are at $\xi>5.5$ where the raw SPTpol candidate list is $>99\%$ pure (Section \ref{subsec:expectedpurity}).  

Comparing the estimated masses we find the median ratio of the masses of common systems in SPTpol/SPT-SZ to be $1.025^{+0.01}_{-0.04}$.  
A plot showing the masses of the common systems within $\delta_z=0.1$ (this cut removes 5 systems at $z > 0.85$) is shown in Figure \ref{fig:mass_comparison}.  
We apply the redshift cut here and in the other SZ sample comparisons in this section to avoid highlighting differences in mass that would purely arise from the masses being estimated at significantly different redshifts. 

\subsubsection{SPTpol 100d}

Next, we compare the SPTpol 500d sample to the sample produced from the SPTpol 100d survey \citep{huang20}. 
This sample consists of 89 clusters detected at $\xi>4.6$.
Of these 89 systems, 86 are in the non-point source masked region and 73 are in the SPTpol 500d sample. 
The matched candidates have a median spatial separation of 0\farcm{22}. 
Using our updated scaling-relation parameters discussed in Section \ref{sec:mass}, we find the median ratio of the masses of clusters in SPTpol 500d/SPTpol 100d to be $1.02\pm 0.02$ for systems within $\delta_z=0.1$.  
This comparison was conducted using 66 systems, which excludes from the 73 matches above 1 unconfirmed candidate, 2 newly confirmed systems in this work,  and 4 systems with updated \spitzer \ redshifts that shifted them outside this range. 
We plot the SPTpol 500d and 100d masses against each other in the middle panel of Figure \ref{fig:mass_comparison}. 
 The $\xi$ values of unmatched SPTpol 100d cluster candidates (8/13 confirmed with redshifts) range from 4.6 to 6.2.

\subsubsection{ACT}
Finally, repeating the matching exercise with the ACT cluster catalog of \citet{hilton21}, we find 141 ACT cluster candidates detected at ACT signal-to-noise $>4$ fall within the SPTpol 500d footprint and 132 in the non-masked region. 
Using the same $3\arcmin$ matching radius as for SPT-SZ, we find 108 of these candidates have matches in the SPTpol sample, with a median ratio of SPTpol/ACT detection signal-to-noise of 1.66 and median spatial separation of 
0$\farcm{33}$.
There are 24 candidates not matched, ranging in ACT detection significance from 4.1 to 5.4$\sigma$ and having a median reported mass of {$M_{500c}=2\times10^{14}$ \msun}. 

Reversing the question, there are 389 SPTpol candidates at $\xi>4$  and north of $\delta=-60$ (the southernmost extent of the ACT sample) not found in the ACT sample, with 290 having been confirmed by optical/IR follow-up observations.
The  $\xi$ values of these unmatched confirmed clusters lie in the range $4<\xi<9$ with a median mass of {$M_{500c}=2.4\times10^{14}$ \msun} .  
Of these systems, 64 are at $\xi>5.5$, a significance above which candidates are highly pure, independent of optical follow-up (see Figure \ref{fig:purity_fig}). 

A  more detailed comparison of the relative completeness of the ACT and SPTpol samples will require careful investigations of the cluster confirmation procedure  as well as masking and spatial variations in the noise levels in the ACT maps.
\citet{hilton21} also did not report a full candidate list, rather only  cluster candidates confirmed by their optical/infrared analysis, which further complicates this comparison. 
The dominant source of follow-up confirmation and redshifts for ACT clusters in the SPTpol region was the redMaPPer algorithm run on DES data  (which has excellent agreement with MCMF at $z<1$, see Figure \ref{fig:redshift_comparison}), followed by the zCluster algorithm run on DESI Legacy Survey data \citep{dey19} which includes optical \textit{grz} photometry combined with WISE {\sl W1} and {\sl W2} channel data, and then previously reported cluster redshifts from SPT and the literature.  
Given the challenges of confirming high-redshift clusters, we would naturally expect most variation in the confirmation fraction of this portion of the sample.

Moving on from the identification of systems, we now compare the estimated masses. 
The ACT collaboration uses a different procedure than SPT for estimating the masses of clusters. 
Following  initial cluster identification, a central Compton-$y$ parameter, $y_o$, is measured for each candidate using a 2\farcm{4} matched filter.
This central value is then corrected for the mismatch as a function of cluster mass and redshift between the expected cluster profile and that adopted for the matched filter. 
The \citet{arnaud10} pressure-mass scaling relation,  statistical and assumed intrinsic scatter in $y_o$, and the \citet{tinker08} mass function (computed at the same fiducial cosmology we adopt in this work) are then used to compute the probability distribution, P(\mass $|$ $y_o$,$z$). 
The maximum in this distribution is assigned as the cluster mass. 
This mass is further corrected by a value of $0.71\pm 0.07$ using the weak lensing $\lambda$- mass calibration of \citet{mcclintock19}; these masses are reported as {$M^\textrm{Cal}_{\mbox{\scriptsize 500c}}$ in the ACT catalog. 

 A plot showing the masses of the common SPTpol 500d/ACT systems within $\delta_z=0.1$ (100 clusters, 8 $z>0.96$ systems excluded for redshift differences) is shown in Figure \ref{fig:mass_comparison}. 
Comparing these masses, we find the ratio of SPTpol/ACT masses to be  $1.07\pm 0.03$.  
We find this ratio to be consistent with the results comparing the median mass ratio of 362 common clusters between SPT-SZ and ACT ($1.06^{+0.02}_{-0.006}$).   
This difference in normalization is larger than the $1.027\pm0.012$ reported in \citet{hilton21}.
That work measured this ratio using 228 common clusters with ACT signal-to-noise $>6$  and the full SPT-SZ sample at $\xi>4.5$; we recover the same mass ratio when applying ACT's thresholds. 
We note given the significant differences in how the mass-scaling relation parameters are determined---SPT's via the abundance matching of our sample to a fiducial cosmology versus ACT's adoption of a weak lensing calibration via cross matching with the DES redMaPPER calibration (the latter scaling known to be biased by projection effects, especially at lower masses \citealt[]{bleem20,grandis21})--- that the agreement here between SPT and ACT masses is not a strong cross check of the mass estimation techniques.

\begin{figure}
\begin{center}
\includegraphics[width=3.5in]{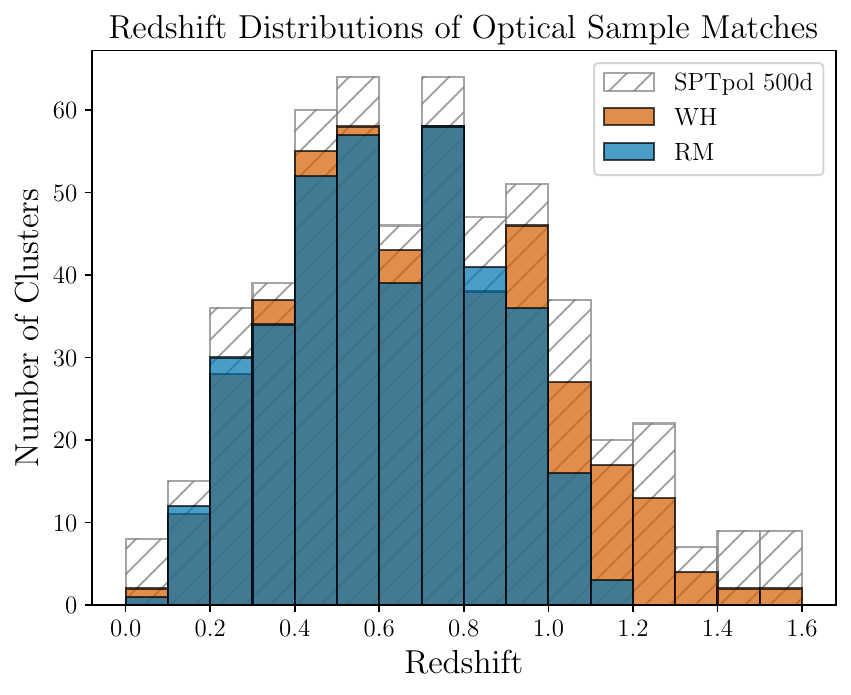}
\caption{Redshift distribution of optical/IR cluster sample matches from RM and WH to the full SPTpol 500d sample. Matches are plotted at the SPTpol redshifts. Overall, the RM sample recovers 70\% of the SPTpol sample and the WH sample recovers 81\%;  many of the lower-redshift clusters not recovered in RM are near regions masked in their analysis.}
\label{fig:opticalmatch}
\end{center}
\end{figure}

\subsubsection{Optically Selected Cluster Samples}\label{subsec:opticalmatch}
 
 We compare the SPTpol 500d sample to two optical samples produced by algorithms that adopt complementary approaches to finding clusters in photometric data. 
The first sample, the DES redMaPPer\footnote{Version 6.4.22+2} sample (hereafter RM), was constructed by running the RM algorithm \citep{rykoff16} on DES Y3 data  \citep{abbott18b}. 
RM identifies clusters by locating spatial over-densities of red-sequence galaxies.  
The full Y3 RM sample consists of 869,335 systems with a weighted galaxy-count mass proxy $\lambda>5$ and redshifts 0.1 $<z<0.95$. 
The second sample, produced by  \citet[][hereafter WH]{wen22}, was constructed using DES and WISE \citep{wright10} data; 
clusters in this sample were  identified as over-densities of stellar mass in photometric redshift slices. 
This sample contains 151,244 clusters from $0.1<z<1.5$ above a detection signal-to-noise of 5. The inclusion of IR data from WISE greatly extends the redshift reach of the catalog compared to that of RM. 

Here, as the spatial density of the optical systems is much greater than that of the SZ surveys, we define a cluster-match between samples as having a counterpart within 2$\arcmin$ and within $\delta_z=0.1\times(1+z)$. When multiple systems are found, we adopt the closest match in redshift as the cluster counterpart. 
In total we match 379/\nconfirm \ (70\%) of the SZ clusters using RM and 441/\nconfirm \ (81\%) using WH. 
We plot the redshift distribution of the SPTpol 500d sample and its matched counterparts in the wide-field optical/IR cluster searches in Figure \ref{fig:opticalmatch}. 

We next compare the redshift consistency for estimates from the different samples.  
We relax our matching criteria to include optical systems matched within 2\arcmin \ for candidates which did not have a good spatial and redshift match to allow us to check for instances where RM and WH may have detected clusters not already captured in our analysis or identified the same optical system at a discrepant redshift.  
To account for the fact that there are candidates for which we have identified lower-significance galaxy excesses along the line-of-sight below our optical confirmation threshold, we assign these unconfirmed systems the redshifts of these over-densities for this exercise only. 
Finally, it is important to remember that the redshift estimates we are comparing here were conducted using similar or identical datasets (DES Years 3 or 6, with WISE for some a subset of the high-$z$ clusters), though with different methods (photo-$z$ versus red-sequence) and with different spectroscopic calibration samples. 

\begin{figure}
\begin{center}
\includegraphics[width=3.5in]{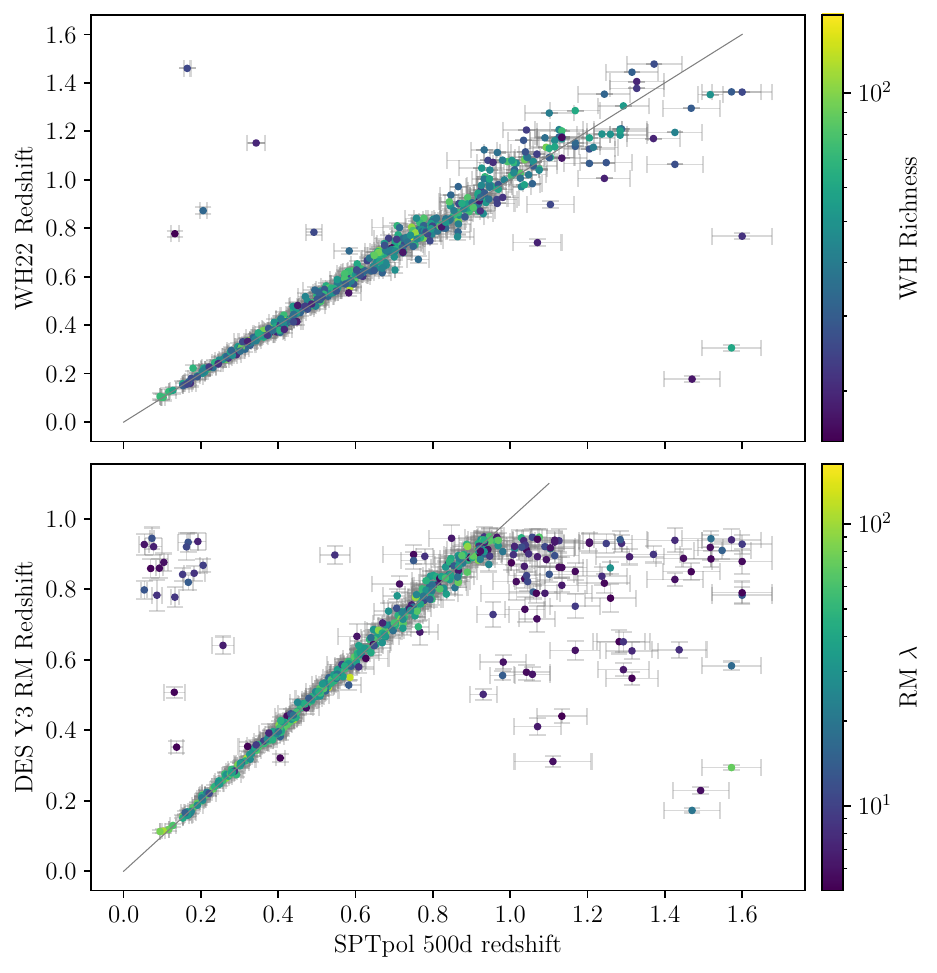}
\caption{Comparison of redshifts of systems from wide-field cluster searches associated with SPTpol 500d candidates by a (1) redshift/spatial proximity match or (2) only a  spatial-proximity match if no systems fell within the redshift association cut (for more details of the matching process see Section \ref{subsec:opticalmatch}). In the  \textit{top} panel are shown matches to the \citet{wen22} optical/IR sample and in the \textit{bottom} panel to the DES RM sample constructed using the DES Y3 dataset. Note that redshift uncertainties are not provided at $z>1$ for the WH sample. Plotted in gray on each panel is a line representing a one-to-one
relationship. The WH (RM) sample has good spatial and redshift matches for 81\% (70\%) of the SPTpol sample. Systems with redshifts that significantly differ between SPTpol and the wide-field optical/IR  searches have typically low wide-field survey richnesses, and are likely spurious associations with the SZ detections. This highlights the importance of accounting for chance associations when matching galaxy over-densities to SZ detections. }
\label{fig:redshift_comparison}
\end{center}
\end{figure}

As shown in Figure \ref{fig:redshift_comparison}, overall there is good agreement in the measured redshifts for matching systems. We find $\sigma_{\delta_z}/(1+z) = 0.009$ (0.014) for RM (WH) at $z<1$ for clusters we deem a spatial/redshift match, with the blind RM search systematically underestimating matched redshifts at $z>0.9$ (beyond its stated redshift reliability).  
At $z>1$, the scatter broadens for the difference between SPTpol 500d and WH to $\sigma_{\delta_z}/(1+z) =0.04$. 
Given their small number of high-$z$ spectroscopic calibrators, WH do not report redshift uncertainties for these highest-$z$ systems. 
The optical/IR systems from the wide-field searches that have discrepant redshifts are typically low-richness and thus likely spurious associations; such discrepant associations are particularly noticeable for \mbox{high-$z$} SPT clusters at redshifts higher than the RM search.
This highlights the importance of quantifying such false association probabilities \citep{klein18,bleem20,hilton21} when mapping associations between optical galaxy over-densities and SZ clusters.

\section{Conclusions}\label{sec:conclusion}

In this work we described  the production of a new sample of galaxy clusters detected by their SZ signature in the SPTpol 500d survey.
We implemented several important  improvements to our cluster catalog production and characterization processes. 
These improvements include the subtraction of moderate signal-to-noise emissive sources from the temperature maps before cluster finding to reduce the number of spurious candidates, improved cluster simulations and sample purity estimation techniques, and new tests to identify potential biases to our recovered SZ signal.

The new SPTpol 500d sample consists of \ncand \ galaxy cluster candidates detected at $\xi>4$.
These candidates were selected from coadded maps of depths 5.3 (11.7) $\mu$K$_{\textrm{CMB}}$-arcmin at 150 (95) GHz that were created from  $>4000$ observations of the SPTpol 500d field. 
To  confirm candidates as clusters, we search optical and IR observations from  DECam, WISE, and  \spitzer  \ with the MCMF algorithm and a related IR-based code to identify significant red-sequence and/or IR galaxy over-densities at the candidate locations.
We  probabilistically confirm \nconfirm \ of these candidates as galaxy clusters, with an expected confirmed sample purity of $\sim94\%$.
The SPTpol 500d cluster sample has a median mass of $2.5 \times 10^{14} $\msun, a median redshift of $z=0.7$,  and is expected to be $>90\%$ complete at $M_{500c} >3.5\times 10^{14}$ \msun \ at $z>0.25$.  
A significant fraction of the systems are at high $z$, with 114 clusters at $z > 1$. 

Masses are estimated for confirmed clusters via a $\xi-$mass scaling relation whose parameters are determined by matching the abundance of observed clusters to a fixed $\Lambda$CDM cosmology. 
Using simulations to relatively calibrate this relation for three independently extracted cluster catalogs from data of variable survey depth---the SPT-SZ \citep{bleem15b}, SPTpol 100d \citep{huang20}, and SPTpol 500d cluster samples---we find excellent agreement for the masses of clusters in common between SPTpol 500d and the other surveys.
We have used both external datasets and internal checks to test for contamination of the SZ signal that could bias these mass estimates.
These investigations include an assessment of potential radio contamination via extrapolating data from the 843 MHz Sydney University Molonglo Sky Survey at cluster locations. This analysis predicts a median contamination of $\Delta \xi_\mathrm{med} = 0.032$ and that  $\sim7\%$ of candidates could have a predicted contamination of greater than $\Delta \xi=1$. 
However, this prediction requires a significant extrapolation from low-frequency data to the SPT bands, and may be an overestimate.
Internal tests comparing masses for clusters detected at  95 and 150 GHz alone show---with the exception of a small number of clusters---insignificant radio or CIB contamination to the SZ signal. 

Matching to previous high-resolution SZ surveys in the region, we find 73, 97, and 141 clusters in common with the SPTpol 100d, SPT-SZ, and ACT \citep{hilton21} cluster samples, respectively.  
These clusters are detected at $\sim1.7\times$ higher signal-to-noise in SPTpol data than SPT-SZ or ACT data. 
We find good agreement between the masses of systems in common between the SPT surveys and that the SPT masses are $\sim$1.07$\times$ higher than the ACT reported masses for the common clusters. 
A spatial-redshift match to the wide-field optical/IR cluster samples of DES Y3 redMaPPer \citep{rykoff16} and the sample produced by \citet{wen22} finds matches for 70-81\% of the SPTpol sample, but also highlights the importance of probabilistic confirmation techniques to avoid spurious associations.

While this is the final cluster sample from the SPTpol collaboration, a number of ongoing astrophysical and cosmological studies are underway that take advantage of its unique low-mass and high-redshift capabilities. 
Although surveys like SPT-3G  \citep{sobrin18}, Simons Observatory \citep{simonsobservatorycollab19}, and CMB-S4   \citep{abazajian19} will in the future surpass the SPTpol survey  with deeper and wider data sets, the work here---particularly the explorations of emissive sources, simulation modeling of SZ cluster samples, and internal and external contamination tests---pave the way for maximizing the potential of the SPTpol and these future surveys for SZ cluster science.

\paragraph{As a final note}
We also release the coadded mm-wave maps and associated data products used to produce this sample. 
The  maps and supporting products are available  at \url{https://pole.uchicago.edu/public/data/sptpol_500d_clusters/index.html}, and the NASA LAMBDA website.
An interactive sky server with the SPTpol maps and Dark Energy Survey data release 2 images is also available at NCSA \url{https://skyviewer.ncsa.illinois.edu}.

\section*{Acknowledgements} 
The South Pole Telescope program is supported by
the National Science Foundation (NSF) through award
OPP-1852617. Partial support is also
provided by the Kavli Institute of Cosmological Physics
at the University of Chicago. 
Work at Argonne National Lab is supported by UChicago Argonne LLC, Operator of Argonne National Laboratory (Argonne). Argonne, a U.S. Department of Energy Office of Science Laboratory, is operated under contract no. DE-AC02-06CH11357. 
This research used resources of the Argonne Leadership Computing Facility, which is supported by DOE/SC under
contract DE-AC02-06CH11357.
TS acknowledges support from the German Federal
Ministry for Economic Affairs and Energy (BMWi) provided
through DLR under projects 50OR2002 and 50OR2302, from the German Research Foundation (DFG) under grant 415537506, and the Austrian Research Promotion Agency (FFG) and the Federal Ministry of the Republic of
Austria for Climate Action, Environment, Mobility, Innovation and
Technology (BMK) via grants 899537 and 900565.
The Melbourne group acknowledges support from the Australian Research Council’s Discovery Projects scheme (No. DP200101068).
This work is based in part on observations made with the Spitzer Space Telescope, which was operated by the Jet Propulsion Laboratory, California Institute of Technology under a contract with NASA.

Funding for the DES Projects has been provided by the U.S. Department of Energy, the U.S. National Science Foundation, the Ministry of Science and Education of Spain, 
the Science and Technology Facilities Council of the United Kingdom, the Higher Education Funding Council for England, the National Center for Supercomputing 
Applications at the University of Illinois at Urbana-Champaign, the Kavli Institute of Cosmological Physics at the University of Chicago, 
the Center for Cosmology and Astro-Particle Physics at the Ohio State University,
the Mitchell Institute for Fundamental Physics and Astronomy at Texas A\&M University, Financiadora de Estudos e Projetos, 
Funda{\c c}{\~a}o Carlos Chagas Filho de Amparo {\`a} Pesquisa do Estado do Rio de Janeiro, Conselho Nacional de Desenvolvimento Cient{\'i}fico e Tecnol{\'o}gico and 
the Minist{\'e}rio da Ci{\^e}ncia, Tecnologia e Inova{\c c}{\~a}o, the Deutsche Forschungsgemeinschaft and the Collaborating Institutions in the Dark Energy Survey. 

The Collaborating Institutions are Argonne National Laboratory, the University of California at Santa Cruz, the University of Cambridge, Centro de Investigaciones Energ{\'e}ticas, 
Medioambientales y Tecnol{\'o}gicas-Madrid, the University of Chicago, University College London, the DES-Brazil Consortium, the University of Edinburgh, 
the Eidgen{\"o}ssische Technische Hochschule (ETH) Z{\"u}rich, 
Fermi National Accelerator Laboratory, the University of Illinois at Urbana-Champaign, the Institut de Ci{\`e}ncies de l'Espai (IEEC/CSIC), 
the Institut de F{\'i}sica d'Altes Energies, Lawrence Berkeley National Laboratory, the Ludwig-Maximilians Universit{\"a}t M{\"u}nchen and the associated Excellence Cluster Universe, 
the University of Michigan, the National Optical Astronomy Observatory, the University of Nottingham, The Ohio State University, the University of Pennsylvania, the University of Portsmouth, 
SLAC National Accelerator Laboratory, Stanford University, the University of Sussex, Texas A\&M University, and the OzDES Membership Consortium.

Based in part on observations at Cerro Tololo Inter-American Observatory, National Optical Astronomy Observatory, which is operated by the Association of 
Universities for Research in Astronomy (AURA) under a cooperative agreement with the National Science Foundation.

The DES data management system is supported by the National Science Foundation under Grant Numbers AST-1138766 and AST-1536171.
The DES participants from Spanish institutions are partially supported by MINECO under grants AYA2015-71825, ESP2015-66861, FPA2015-68048, SEV-2016-0588, SEV-2016-0597, and MDM-2015-0509, 
some of which include ERDF funds from the European Union. IFAE is partially funded by the CERCA program of the Generalitat de Catalunya.
Research leading to these results has received funding from the European Research
Council under the European Union's Seventh Framework Program (FP7/2007-2013) including ERC grant agreements 240672, 291329, and 306478.
We  acknowledge support from the Brazilian Instituto Nacional de Ci\^encia
e Tecnologia (INCT) e-Universe (CNPq grant 465376/2014-2).

This manuscript has been authored by Fermi Research Alliance, LLC under Contract No. DE-AC02-07CH11359 with the U.S. Department of Energy, Office of Science, Office of High Energy Physics. The United States Government retains and the publisher, by accepting the article for publication, acknowledges that the United States Government retains a non-exclusive, paid-up, irrevocable, world-wide license to publish or reproduce the published form of this manuscript, or allow others to do so, for United States Government purposes.

This research has made use of the VizieR catalogue access tool, CDS,
 Strasbourg, France (DOI : 10.26093/cds/vizier). The original description 
 of the VizieR service was published in 2000, A\&AS 143, 23.
 
 This research made use of APLpy, an open-source plotting package for Python \citep{aplpy2012, aplpy2019}.

\textit{Facilities:} Blanco (DECAM), NSF/US Department of Energy, 10m South Pole Telescope (SPTpol), Spitzer (IRAC), WISE}

\bibliographystyle{yahapj}
\bibliography{./spt}

\clearpage
\newpage
\mbox{~}
\clearpage
\newpage

\LongTables

\begin{deluxetable*}{ lcc | cc |c|c|cc|c}
\tablecaption{Galaxy Cluster Candidates at $\xi>4$ in the SPTpol 500d Survey
\label{tab:candidates}}
\tablehead{
\multicolumn{1}{c}{\bf SPT ID} & 
\multicolumn{1}{c}{\bf R.A.} & 
\multicolumn{1}{c}{\bf Decl.} & 
\multicolumn{2}{c}{\bf Best} &
\multicolumn{1}{c}{\bf Redshift} &
\multicolumn{1}{c}{\bf M$_{500c}$}  &
\multicolumn{1}{c}{\bf $\lambda$}  &
\multicolumn{1}{c}{\bf \textit{Spitzer}} &
\multicolumn{1}{c}{\bf Notes} \\
\colhead{} & 
\colhead{(J2000)} & 
\colhead{(J2000)}  & 
\colhead{\nSN{}} & 
\colhead{$\theta_\textrm{c}$} & 
\colhead{} &
\colhead{($10^{14} h_{70}^{-1} M_\odot$)} &
\colhead{}  &
\colhead{}  &
\colhead{}
}
\startdata 
\input{catalog_tablevOct3}
\enddata
\tablecomments{Galaxy cluster candidates detected at $\xi>4$ in the SPTpol 500d survey. We first report properties derived solely from the SPT data, namely the candidate names, positions, detection-significances ($\xi$) and corresponding  $\beta$-model core-radii at which the candidates were detected. Next, using the optical/IR confirmation processes discussed in Section \ref{sec:confirmation}, we provide redshifts of associated galaxy counterparts to the SZ detections for systems with $f_\textrm{cont}<0.2$. These redshifts are combined with the $\xi$ values to estimate cluster masses (Section \ref{sec:mass}) for systems with optical/IR contamination fractions such that the overall sample is expected to be 94\% pure. We provide the optical richness ($\lambda$) and mark whether these quantities were derived with our MCMF or \spitzer \ analysis.  Finally we flag \textbf{LL} confirmed clusters with redshift lower-limits of $z=1.6$ (see Section \ref{subsec:spitzer}),  \textbf{RC} clusters in which potentially $\Delta\xi>2$ is filled in by synchrotron emission (see Section \ref{sec:biases}) or \textbf{LOS} for which a secondary optical structure along the line-of-sight met our confirmation criteria. The richnesses, redshifts, and contamination fractions of both primary and secondary systems are reported in the online version of the cluster sample.  We also provide in the online file information on line-of-sight structures for unconfirmed clusters.}
\end{deluxetable*}

\end{document}

%% file: 500d_authors.tex
\def\ANLHEP{1}
\def\KICPChicago{2}
\def\LMUO{3}
\def\MPE{4}
\def\CTIO{5}
\def\Cardiff{6}
\def\Linea{7}
\def\Michigan{8}
\def\FNAL{9}
\def\Melbourne{10}
\def\ILAst{11}
\def\ILNCSA{12}
\def\CfAb{13}
\def\NIST{14}
\def\ColoradoPhys{15}
\def\Portsmouth{16}
\def\AAUChicago{17}
\def\KIPAC{18}
\def\Stanford{19}
\def\SLAC{20}
\def\UCLondon{21}
\def\MIT{22}
\def\PhysicsUChicago{23}
\def\EFIChicago{24}
\def\Canarias{25}
\def\Laguna{26}
\def\IFAE{27}
\def\McGill{28}
\def\UKZN{29}
\def\UChicago{30}
\def\JPL{31}
\def\Trieste{32}
\def\INAF{33}
\def\IFPU{34}
\def\Cornell{35}
\def\IPNS{36}
\def\KEK{37}
\def\CIEMAT{38}
\def\IITHyderabad{39}
\def\CIFAR{40}
\def\ColoradoAPS{41}
\def\Oslo{42}
\def\Miss{43}
\def\HarveyMudd{44}
\def\UAM{45}
\def\Penn{46}
\def\esogarching{47}
\def\Innsbruck{48}
\def\Queensland{49}
\def\ILPhys{50}
\def\SantaCruz{51}
\def\Berkeley{52}
\def\CCAPP{53}
\def\OSU{54}
\def\Grenoble{55}
\def\CfA{56}
\def\Davis{57}
\def\AAO{58}
\def\Lowell{59}
\def\LBNL{60}
\def\TexasAM{61}
\def\Catalana{62}
\def\ExcellenceCluster{63}
\def\AstroPrinceton{64}
\def\Dunlap{65}
\def\UToronto{66}
\def\ANLMSD{67}
\def\Rio{68}
\def\Caltech{69}
\def\Hamburg{70}
\def\Minnesota{71}
\def\CNRS{72}
\def\Sussex{73}
\def\CaseWestern{74}
\def\Brook{75}
\def\Saclay{76}
\def\ArtInstChicago{77}
\def\Bonn{78}
\def\ThreeSpeedLogic{79}
\def\Southampton{80}
\def\Chulalongkorn{81}
\def\Rubin{82}
\def\OakRidge{83}
\def\Southwest{84}
\def\MSU{85}

\author{
  L.~E.~Bleem\altaffilmark{\ANLHEP,\KICPChicago},
  M.~Klein\altaffilmark{\LMUO,\MPE},
  T.~M.~C.~Abbott\altaffilmark{\CTIO},
  P.~A.~R.~Ade\altaffilmark{\Cardiff},
  M.~Aguena\altaffilmark{\Linea},
  O.~Alves\altaffilmark{\Michigan},
  A.~J.~Anderson\altaffilmark{\FNAL},
  F.~Andrade-Oliveira\altaffilmark{\Michigan},
  B.~Ansarinejad\altaffilmark{\Melbourne},
  M.~Archipley\altaffilmark{\ILAst,\ILNCSA}
% ...
}

\altaffiltext{\ANLHEP}{High-Energy Physics Division, Argonne National Laboratory, 9700 South Cass Avenue., Lemont, IL, 60439, USA}
\altaffiltext{\KICPChicago}{Kavli Institute for Cosmological Physics, University of Chicago, 5640 South Ellis Avenue, Chicago, IL, 60637, USA}
\altaffiltext{\LMUO}{University Observatory, Faculty of Physics, Ludwig-Maximilians-Universit\"at, Scheinerstr. 1, 81679 Munich, Germany}
\altaffiltext{\MPE}{Max-Planck-Institut f\"{u}r extraterrestrische Physik,Giessenbachstr.\ 85748 Garching, Germany}
\altaffiltext{\CTIO}{Cerro Tololo Inter-American Observatory, NSF's National Optical-Infrared Astronomy Research Laboratory, Casilla 603, La Serena, Chile}
\altaffiltext{\Cardiff}{School of Physics and Astronomy, Cardiff University, Cardiff CF24 3YB, United Kingdom}
\altaffiltext{\Linea}{Laborat\'orio Interinstitucional de e-Astronomia - LIneA, Rua Gal. Jos\'e Cristino 77, Rio de Janeiro, RJ - 20921-400, Brazil}
\altaffiltext{\Michigan}{Department of Physics, University of Michigan, 450 Church Street, Ann Arbor, MI, 48109, USA}
\altaffiltext{\FNAL}{Fermi National Accelerator Laboratory, MS209, P.O. Box 500, Batavia, IL, 60510, USA}
\altaffiltext{\Melbourne}{School of Physics, University of Melbourne, Parkville, VIC 3010, Australia}
\altaffiltext{\ILAst}{Department of Astronomy, University of Illinois Urbana-Champaign, 1002 West Green Street, Urbana, IL, 61801, USA}
\altaffiltext{\ILNCSA}{Center for AstroPhysical Surveys, National Center for Supercomputing Applications, Urbana, IL, 61801, USA}
\altaffiltext{\CfAb}{Center for Astrophysics  $|$ Harvard \& Smithsonian, Optical and Infrared Astronomy Division, Cambridge, MA 01238, USA}
\altaffiltext{\NIST}{NIST Quantum Devices Group, 325 Broadway Mailcode 817.03, Boulder, CO, 80305, USA}
\altaffiltext{\ColoradoPhys}{Department of Physics, University of Colorado, Boulder, CO, 80309, USA}
\altaffiltext{\Portsmouth}{Institute of Cosmology and Gravitation, University of Portsmouth, Portsmouth, PO1 3FX, UK}
\altaffiltext{\AAUChicago}{Department of Astronomy and Astrophysics, University of Chicago, 5640 South Ellis Avenue, Chicago, IL, 60637, USA}
\altaffiltext{\KIPAC}{Kavli Institute for Particle Astrophysics and Cosmology, Stanford University, 452 Lomita Mall, Stanford, CA, 94305, USA}
\altaffiltext{\Stanford}{Department of Physics, Stanford University, 382 Via Pueblo Mall, Stanford, CA, 94305, USA}
\altaffiltext{\SLAC}{SLAC National Accelerator Laboratory, 2575 Sand Hill Road, Menlo Park, CA, 94025, USA}
\altaffiltext{\UCLondon}{Department of Physics \& Astronomy, University College London, Gower Street, London, WC1E 6BT, UK}
\altaffiltext{\MIT}{Kavli Institute for Astrophysics and Space Research, Massachusetts Institute of Technology, 77 Massachusetts Avenue, Cambridge, MA~02139, USA}
\altaffiltext{\PhysicsUChicago}{Department of Physics, University of Chicago, 5640 South Ellis Avenue, Chicago, IL, 60637, USA}
\altaffiltext{\EFIChicago}{Enrico Fermi Institute, University of Chicago, 5640 South Ellis Avenue, Chicago, IL, 60637, USA}
\altaffiltext{\Canarias}{Instituto de Astrofisica de Canarias, E-38205 La Laguna, Tenerife, Spain}
\altaffiltext{\Laguna}{Universidad de La Laguna, Dpto. Astrofísica, E-38206 La Laguna, Tenerife, Spain}
\altaffiltext{\IFAE}{Institut de F\'{\i}sica d'Altes Energies (IFAE), The Barcelona Institute of Science and Technology, Campus UAB, 08193 Bellaterra (Barcelona) Spain}
\altaffiltext{\McGill}{Department of Physics and McGill Space Institute, McGill University, 3600 Rue University, Montreal, Quebec H3A 2T8, Canada}
\altaffiltext{\UKZN}{School of Mathematics, Statistics \& Computer Science, University of KwaZulu-Natal, Durban, South Africa}
\altaffiltext{\UChicago}{University of Chicago, 5640 South Ellis Avenue, Chicago, IL, 60637, USA}
\altaffiltext{\JPL}{Jet Propulsion Laboratory, California Institute of Technology, Pasadena, CA 91011, USA}
\altaffiltext{\Trieste}{Astronomy Unit, Department of Physics, University of Trieste, via Tiepolo 11, I-34131 Trieste, Italy}
\altaffiltext{\INAF}{INAF-Osservatorio Astronomico di Trieste, via G. B. Tiepolo 11, I-34143 Trieste, Italy}
\altaffiltext{\IFPU}{Institute for Fundamental Physics of the Universe, Via Beirut 2, 34014 Trieste, Italy}
\altaffiltext{\Cornell}{Cornell University, Ithaca, NY 14853, USA}
\altaffiltext{\IPNS}{Institute of Particle and Nuclear Studies (IPNS), High Energy Accelerator Research Organization (KEK), Tsukuba, Ibaraki 305-0801, Japan}
\altaffiltext{\KEK}{International Center for Quantum-field Measurement Systems for Studies of the Universe and Particles (QUP-WPI), High Energy Accelerator Research Organization (KEK), Tsukuba, Ibaraki 305-0801, Japann}
\altaffiltext{\CIEMAT}{Centro de Investigaciones Energ\'eticas, Medioambientales y Tecnol\'ogicas (CIEMAT), Madrid, Spain}
\altaffiltext{\IITHyderabad}{Department of Physics, IIT Hyderabad, Kandi, Telangana 502285, India}
\altaffiltext{\CIFAR}{Canadian Institute for Advanced Research, CIFAR Program in Gravity and the Extreme Universe, Toronto, ON, M5G 1Z8, Canada}
\altaffiltext{\ColoradoAPS}{Department of Astrophysical and Planetary Sciences, University of Colorado, Boulder, CO, 80309, USA}
\altaffiltext{\Oslo}{Institute of Theoretical Astrophysics, University of Oslo. P.O. Box 1029 Blindern, NO-0315 Oslo, Norway}
\altaffiltext{\Miss}{Faculty of Physics and Astronomy, University of Missouri--Kansas City, 5110 Rockhill Road, Kansas City, MO 64110, USA}
\altaffiltext{\HarveyMudd}{Harvey Mudd College, 301 Platt Boulevard, Claremont, CA, 91711, USA}
\altaffiltext{\UAM}{Instituto de Fisica Teorica UAM/CSIC, Universidad Autonoma de Madrid, 28049 Madrid, Spain}
\altaffiltext{\Penn}{Department of Physics and Astronomy, University of Pennsylvania, Philadelphia, PA 19104, USA}
\altaffiltext{\esogarching}{European Southern Observatory, Karl-Schwarzschild-Str.,DE-85748 Garching b. Munchen, Germany}
\altaffiltext{\Innsbruck}{Universit\"at Innsbruck, Institut f\"ur Astro- und Teilchenphysik, Technikerstr. 25/8, 6020 Innsbruck, Austria}
\altaffiltext{\Queensland}{School of Mathematics and Physics, University of Queensland,  Brisbane, QLD 4072, Australia}
\altaffiltext{\ILPhys}{Department of Physics, University of Illinois Urbana-Champaign, 1110 West Green Street, Urbana, IL, 61801, USA}
\altaffiltext{\SantaCruz}{Santa Cruz Institute for Particle Physics, Santa Cruz, CA 95064, USA}
\altaffiltext{\Berkeley}{Department of Physics, University of California, Berkeley, CA, 94720, USA}
\altaffiltext{\CCAPP}{Center for Cosmology and Astro-Particle Physics, The Ohio State University, Columbus, OH 43210, USA}
\altaffiltext{\OSU}{Department of Physics, The Ohio State University, Columbus, OH 43210, USA}
\altaffiltext{\Grenoble}{LPSC Grenoble - 53, Avenue des Martyrs 38026 Grenoble, France}
\altaffiltext{\CfA}{Harvard-Smithsonian Center for Astrophysics, 60 Garden Street, Cambridge, MA, 02138, USA}
\altaffiltext{\Davis}{Department of Physics, University of California, One Shields Avenue, Davis, CA, 95616, USA}
\altaffiltext{\AAO}{Australian Astronomical Optics, Macquarie University, North Ryde, NSW 2113, Australia}
\altaffiltext{\Lowell}{Lowell Observatory, 1400 Mars Hill Rd, Flagstaff, AZ 86001, USA}
\altaffiltext{\LBNL}{Physics Division, Lawrence Berkeley National Laboratory, Berkeley, CA, 94720, USA}
\altaffiltext{\TexasAM}{George P. and Cynthia Woods Mitchell Institute for Fundamental Physics and Astronomy, and Department of Physics and Astronomy, Texas A\&M University, College Station, TX 77843,  USA}
\altaffiltext{\Catalana}{Instituci\'o Catalana de Recerca i Estudis Avan\c{c}ats, E-08010 Barcelona, Spain}
\altaffiltext{\ExcellenceCluster}{Excellence Cluster Universe, Boltzmannstr.\ 2, 85748 Garching, Germany}
\altaffiltext{\AstroPrinceton}{Department of Astrophysical Sciences, Princeton University, Peyton Hall, Princeton, NJ 08544, USA}
\altaffiltext{\Dunlap}{Dunlap Institute for Astronomy \& Astrophysics, University of Toronto, 50 St. George Street, Toronto, ON, M5S 3H4, Canada}
\altaffiltext{\UToronto}{David A. Dunlap Department of Astronomy \& Astrophysics, University of Toronto, 50 St. George Street, Toronto, ON, M5S 3H4, Canada}
\altaffiltext{\ANLMSD}{Materials Sciences Division, Argonne National Laboratory, 9700 South Cass Avenue, Lemont, IL, 60439, USA}
\altaffiltext{\Rio}{Observat\'orio Nacional, Rua Gal. Jos\'e Cristino 77, Rio de Janeiro, RJ - 20921-400, Brazil}
\altaffiltext{\Caltech}{California Institute of Technology, 1200 East California Boulevard., Pasadena, CA, 91125, USA}
\altaffiltext{\Hamburg}{Hamburger Sternwarte, Universit\"{a}t Hamburg, Gojenbergsweg 112, 21029 Hamburg, Germany}
\altaffiltext{\Minnesota}{School of Physics and Astronomy, University of Minnesota, 116 Church Street SE Minneapolis, MN, 55455, USA}
\altaffiltext{\CNRS}{Laboratoire de physique des 2 infinis Ir\`ene Joliot-Curie, CNRS Universit\'e Paris-Saclay, B\^at. 100, Facult\'e des sciences, F-91405 Orsay Cedex, France}
\altaffiltext{\Sussex}{Department of Physics and Astronomy, Pevensey Building, University of Sussex, Brighton, BN1 9QH, UK}
\altaffiltext{\CaseWestern}{Department of Physics, Case Western Reserve University, Cleveland, OH, 44106, USA}
\altaffiltext{\Brook}{Brookhaven National Laboratory, Upton, NY 11973, USA}
\altaffiltext{\Saclay}{Universit\'{e} Paris-Saclay, CNRS, Institut d'Astrophysique Spatiale, 91405, Orsay, France}
\altaffiltext{\ArtInstChicago}{Liberal Arts Department, School of the Art Institute of Chicago, 112 South Michigan Avenue, Chicago, IL,60603, USA }
\altaffiltext{\Bonn}{Argelander-Institut f\"ur Astronomie, Auf dem H\"ugel 71, D-53121 Bonn, Germany}
\altaffiltext{\ThreeSpeedLogic}{Three-Speed Logic, Inc., Victoria, B.C., V8S 3Z5, Canada}
\altaffiltext{\Southampton}{School of Physics and Astronomy, University of Southampton,  Southampton, SO17 1BJ, UK}
\altaffiltext{\Chulalongkorn}{Department of Physics, Faculty of Science, Chulalongkorn University, 254 Phayathai Road, Pathumwan, Bangkok Thailand. 10330}
\altaffiltext{\Rubin}{Vera C. Rubin Observatory Project Office, 933 North Cherry Avenue, Tucson, AZ 85721, USA}
\altaffiltext{\OakRidge}{Computer Science and Mathematics Division, Oak Ridge National Laboratory, Oak Ridge, TN 37831}
\altaffiltext{\Southwest}{Space Science and Engineering Division, Southwest Research Institute, San Antonio, TX 78238}
\altaffiltext{\MSU}{Department of Physics and Astronomy, Michigan State University, East Lansing, MI 48824, USA}

%% file: catalog_tablevOct3.tex
SPT-CL J0000$-$6020 & 0.0348 & $-60.3367$ & 7.15 & 0.25 & 0.76$\pm$0.02 & $3.15^{+0.43}_{-0.52}$ & 51$\pm$8 & 0 & -\\
SPT-CL J0000$-$5825 & 0.0855 & $-58.4250$ & 4.03 & 0.50 & 0.74$\pm$0.03 & $2.14^{+0.33}_{-0.47}$ & 30$\pm$7 & 0 & -\\
SPT-CL J0000$-$6301 & 0.1269 & $-63.0190$ & 4.44 & 0.25 & 0.245$\pm$0.008 & $2.67^{+0.44}_{-0.57}$ & 55$\pm$9 & 0 & -\\
SPT-CL J0000$-$6121 & 0.1890 & $-61.3538$ & 4.01 & 0.25 & - & - & - & 0 & -\\
SPT-CL J0000$-$5748 & 0.2479 & $-57.8083$ & 14.58 & 0.25 & 0.702 & $5.12^{+0.56}_{-0.69}$ & 85$\pm$10 & 0 & RC\\
SPT-CL J0001$-$5917 & 0.3674 & $-59.2986$ & 4.09 & 0.25 & 0.57$\pm$0.03 & $2.27^{+0.36}_{-0.49}$ & 17$\pm$5 & 0 & -\\
SPT-CL J0001$-$5552 & 0.3867 & $-55.8667$ & 4.10 & 0.50 & 0.28$\pm$0.01 & $2.51^{+0.41}_{-0.55}$ & 28$\pm$6 & 0 & -\\
SPT-CL J0001$-$6258 & 0.4021 & $-62.9795$ & 4.86 & 0.50 & 0.22$\pm$0.01 & $2.83^{+0.49}_{-0.59}$ & 26$\pm$6 & 0 & -\\